# UNIVERSIDADE DE LISBOA
Faculdade de Ciências
Departamento de Informática

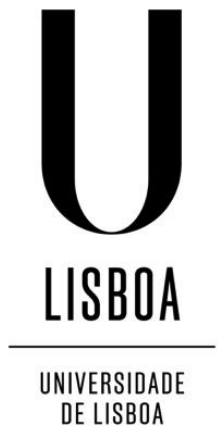

# UCAT
# UBIQUITOUS CONTEXT AWARENESS
# TOOLS FOR THE BLIND

**Ivo Rafael Pereira Rodrigues**

DISSERTAÇÃO

MESTRADO EM ENGENHARIA INFORMÁTICA
Especialização em Sistemas de Informação

2013

UNIVERSIDADE DE LISBOA

Faculdade de Ciências
Departamento de Informática

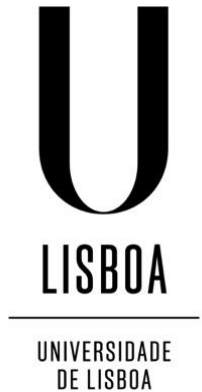

# UCAT
# UBIQUITOUS CONTEXT AWARENESS TOOLS FOR THE BLIND

Ivo Rafael Pereira Rodrigues

DISSERTAÇÃO

Trabalho orientado pelo Prof. Doutor Tiago Guerreiro

MESTRADO EM ENGENHARIA INFORMÁTICA
Especialização em Sistemas de Informação

2013

# Acknowledgements

I would like to thank first and foremost my thesis adviser professor Tiago Guerreiro for accepting to be my mentor for this thesis and enduring with me throughout the year as well as professor Luis Carriço for accepting me and guiding me through the thesis with invaluable input. A mention to Luis Duarte also for the help provided discussing ideas and providing input whenever it was needed.

My family for providing me with the platform and the ability to be able to do this work and have my master degree, without family we can only go so far in life.

Colleagues that through the year shared the journey and always provided a source of knowledge and information when debating the project as well as the much needed moments to clear my head and calm down for a bit when needed.

A thank you also for the FCUL for being a great place to study and grow both on a personal and academic level, the Lasige group who welcomed me during my thesis in particular the Absinth group which provided with, in my opinion, one of the best interactions and participation they could have possibly give for someone who is working on a thesis.

Last but not least friends. These are always the pillars of our stability the people we look for when we need anything, and they were always there whenever I needed them.

# Resumo

As pessoas com deficiências visuais sofrem de uma constante falta de informação sobre o ambiente que as rodeia. Esta informação pode estar relacionada com locais, objetos e até mesmo pessoas. A inexistência de ajuda para combater esta falta de informação faz com que surjam situações em que uma pessoa cega só consegue resolver o seu problema recorrendo a ajuda de outras pessoas. São exemplos a identificação da presença de pessoas, informação escrita o e outros detalhes de uma rotina diária. Os dispositivos móveis dos dias de hoje vêm equipados com uma grande extensão de sensores que possibilitam colmatar algumas destas falhas.

Uma pessoa normovisual está habituada a recorrer à visão para obter informação sobre o ambiente que a rodeia. Sem grande esforço pode obter informação sobre proximidade de pessoas ou a sua movimentação. A mesma facilidade ocorre em reconhecer objetos ou orientar-se e compreender os locais onde se encontra. Esta informação é algo que pessoas cegas têm dificuldade em obter.

Situações normais como identificar quem são as pessoas que os rodeiam torna-se uma tarefa mais complicada especialmente se estas não seguirem as regras de etiqueta social e fizerem notar a sua presença. A acessibilidade nos dispositivos móveis é uma área que ainda está a ser explorada e existem diversas possibilidades de uso que estes dispositivos permitem que ainda não são acessíveis para pessoas cegas.

Cada vez mais os dispositivos móveis, em particular desde a chegada do iPhone, fornece a utilizadores cegos a possibilidade de acederem aplicações sociais, de lazer e de produtividade. Outros sistemas operativos e dispositivos seguem pelo mesmo caminho. O conjunto de características disponibilizadas por estes dispositivos nomeadamente para a localização, comunicação, armazenamento e processamento permitem criar ferramentas de perceção do ambiente para cegos cada vez mais ricas.

Existem diversos trabalhos realizados para melhorar a orientação, localização, reconhecimento de objectos e proximidade (na maioria dos casos a obstáculos). Muitos deles focam-se em ambientes interiores, outros em ambientes exteriores. Uns em objects, outros têm uma componente maos social. O problema de todos estes projectos é que nenhum deles oferece uma solução complete e acessivel para pessoas cegas. As falhas existentes em cada um destes projectos deixa em aberto a possibilidade para que fosse desenvolvido um Sistema que não só junta-se todas as caracteristicas de awareness que pudessem ajudar uma pessoa cega mas também tornar este Sistema acessivel.

No entanto mesmo com tantas aplicações e projectos existe uma falha no que toca a ferramentas que tratem de perceção implícita, que forneça informação sobre o ambiente que nos rodeia dentro de contexto. O nosso projeto focou-se em identificar as necessidades de informação de uma pessoa cega no que diz respeito à falta de informação.

Foi realizada uma entrevista preliminar para que pudéssemos perceber as limitações e as necessidades que as pessoas cegas enfrentam quando confrontadas com ambientes sociais. A entrevista tinha como foco um perfil básico dos utilizadores e questões sobre o seu uso da tecnologia utilizada no projeto assim como questões sobre cenários onde a perceção do ambiente é limitada.

A grande maioria da entrevista focou-se em questões sobre perceção, particularmente tendo em atenção a orientação em ambientes novos e ambientes já conhecidos, as dificuldades que normalmente encontram em situações sociais e de que maneira ou que comportamentos desenvolveram para tentar facilitar o seu dia-a-dia. Tentamos perceber quais as maiores fontes de desconforto que estão habituados a enfrentar e quais delas é que causam mais transtorno.

A primeira causa apontada como fonte de desconforto foi a falta de conhecimento das pessoas que os rodeiam, tanto saber quem são como quantas pessoas são. Também fizemos a questão sobre o quão fácil é para eles ter a perceção de quem os rodeia, se é uma tarefa fácil de realizar e especialmente se tinham dificuldade em perceber a entrada e saída de pessoas no seu espaço.

Todos os participantes afirmaram que seria bom obter mais informação sobre o ambiente que os rodeia, apesar de apontarem que esta ferramenta deveria passar despercebida e ser subtil na sua tarefa.

Após analisarmos os resultados das entrevistas e de termos em conta os cenários mencionados pelos utilizadores assim como os cenários que tínhamos pensado para a implementação do sistema foi desenvolvido um sistema com a intenção de explorar estes novos cenários e fazer uma avaliação mais completa da aplicação e os seus benefícios.

Após a implementação do sistema este foi submetido a uma fase de avaliação que durou uma semana e meia efetuada com utilizadores cegos para que pudéssemos tirar resultados não só do resultado da implementação do sistema mas do seu impacto para um utilizador cego.

Sobre este Sistema foi feita uma avaliação de uma semana e meia, efectuada com utilizadores cegos durante a qual eles tiveram a oportunidade de explorer as capacidades do Sistema e utilizar durante o seu dia a dia normal a aplicaçao e retirar o máximo que conseguissem. Foi uma avaliação sem guião, logo os utilizadores não tinham tarefas explicitas apenas tinham que utilizar o Sistema à sua conveniencia.

Durante esse period de avaliaçao os utilizadores deram-nos feedback diário de como se estava a comportar o Sistema para que pudessemos obter dados de uso diário. Foi também efectuado no final da avaliação questionários e uma entrevista final para recolher todas as informaçoes possiveis por parte dos utilizadores que fizeram parte da avaliação. Analisamos esses dados e retirámos algumas conclusões.

No geral os utilizadores estiveram satisfeitos com a informação produzida pela aplicação, foi de facto um aumento, na grande maioria dos casos, aquilo a que estão normalmente habituados. Sempre que eram notificados sentiram necessidade de explorar a notificação e perceber melhor qual a informação que a aplicação lhes estava a tentar fornecer.

Verificou-se o interesse por parte dos participantes quando estes tentavam procurar por pessoas novas implicitamente.

Todos os participantes confirmaram que o conhecimento sobre o ambiente que os rodeava foi de facto enriquecido. Houve mais facilidade em identificar pessoas quando estas chegavam a um local.

Nos últimos anos tem sido feito um esforço para aumentar a acessibilidade a dispositivos e à informação disponível nestes (Ex: através de leitores de ecrã). No entanto, o mundo real apresenta muita informação que é oferecida de forma visualmente que é assim inacessível a uma pessoa cega. Esta é uma limitação a ter em consideração que pode ajudar à interação social e a compreensão do ambiente que rodeia o utilizador.

Não existem sistemas que sejam capazes de fornecer a uma pessoa cega informação tão simples como quem é que se encontra à sua volta, quem é que passou próximo de si, quem é que ainda se encontra numa sala, quais as lojas mais próximas de si ou restaurantes ou simplesmente o que está escrito no placard de notícias. Da mesma maneira, existe a falta de ferramentas que lhes permitam adicionar a sua própria informação a ambientes, que possam partilhar essa informação e servirem-se dela para se entreajudarem em situações mais complicadas.

Mostramos estas limitações, necessidades e desejos das pessoas cegas em obter informação sobre o ambiente que as rodeia. Tentando dar o enfâse no aspeto social da ferramenta, focámo-nos muito nas pessoas e na necessidade pessoal de cada individuo

# Abstract

Visually impaired people are often confronted with new environments and they find themselves face to face with an innumerous amount of difficulties when facing these environments. Having to surpass and deal with these difficulties that arise with their condition is something that we can help diminish. They are one sense down when trying to understand their surrounding environments and gather information about what is happening around them. It is difficult for blind people to be comfortable in places where they can't achieve a proper perception of the environment, considering the difficulty to understand where they are, where are the things they want, who are the people around them, what is around them and how to safely get somewhere or accomplish a particular task.

Nowadays, mobile devices present significant computing and technological capacity which has been increasing to the point where it is very common for most people to have access to a device with Bluetooth, GPS, Wi-Fi, and both high processing and storage capacities. This allows us to think of applications that can do so much to help people with difficulties. In the particular case of blind people, the lack of visual information can be bypassed with other contextual information retrieved by their own personal devices.

Mobile devices are ubiquitous and are able to be used virtually anywhere and allow connectivity with one another. This also allows their users to save information and convey it through several devices, which means that we can easily share data, augmenting our possibilities beyond what a single device can do.

Our goal is to provide information to blind users, be able to give them information about the context that surrounds them. We wanted to provide the blind users with the tools to create information and be able to share this information between each other, information about people, locations or objects. Our approach was to split the project into a data and information gathering phase where we did our field search and interviewed and elaborated on how is the situation of environment perception for blind users, followed by a technical phase where we implement a system based on the first stage. Our results gathered from both the collecting phase and our implementing phase showed that there is potential to use these tools in the blind community and that they welcome the possibilities and horizons that it opens them.

**Keywords: Context-Awareness, Blind, Accessibility, Mobile Computing**

# Content

# List of Figures

# List of tables

# Chapter 1 - Introduction

## 1.1 - Motivation

The number of people with visual disabilities worldwide is around 135 million, of which 45 million are blind[1]. This amounts to a significant number of people whose lives are affected by other millions of simple, but at the same time impairing problems due to their condition.

Blind people have a significantly higher difficulty in exploring and being able to know their surroundings, both in indoor and outdoor environments [10]. These difficulties deter and sometimes force blind people from doing some activities they would otherwise enjoy doing or would like to do.

It is common for people to have to go through several places both indoor and outdoor on a daily basis. From their own homes to the work place, public places, street and even recreational spaces. This is no exception for the blind. – However, in their case, the level of comfort or how easy they can orient themselves in these places is much different. For instance, when they move into a hometown, a work place or a school, these people will have difficulties getting to where they want, having a perception of their environment and how to freely do what they want. This is hard already even when they are used to the places, now think about new environments, getting to know these comes with a higher degree of difficulty then usual to them and it even requires them to learn specific information about the environment that non-visually impaired people are not aware of. This knowledge is what they need in order to make their interaction in this environment easier. Due to the fact that we are used to seeing we don't usually take into consideration the kind of information that visually impaired people need to have to be comfortable in a new environment. We thought about how visually impaired people adapt and adjust to a environments especially an indoor environment and we realized that they could not understand when message boards, specific locations (specifically in new places) or even other people had some information that could benefit them or help them. In a world where so much of what we do is dictated by how much we know of what surrounds us, visually impaired people face a great disadvantage that can be improved by trying to provide not only information that non impaired users are used to receive through vision but also provide them with the kind of specialized information that they need because of their condition and also any kind of information that even sighted users get from some tools that simply are not yet accessible to the blind population.

---

[1] Unite for Sight, World Health Organization report, http://www.uniteforsight.org/eye_stats.php), 2009.

The past decade had several technological advances that have made possible the proliferation of personal location technology in portable devices (Smartphones, Tables). Smartphone sales showed strong growth worldwide in 2011. Here are some numbers from several sources[2], report from IDC, (February 2012), says the shipments in 2011 were 491.4 million units up 61.3 percent from 2010. This makes smartphones 31.8 percent of all handsets shipped. Report from Strategy Analytics (February 2012) claims total shipments in 2011 were 488.5 million units up 63.1 percent from 2010. This makes smartphones 31.5 percent of all handsets shipped. Also Gartner (February 2012) reports a total of smartphone sales in 2011 reached 472 million units up 58 percent from 2010. This makes smartphones 31 percent of all handsets shipped. The last IDC report (June 2012) predicts that 686 million smartphones will be sold in 2012, 38.4 percent of all handsets shipped. According to IDC, smart phones accounted for 36 percent of global mobile-phone shipments in the first quarter of 2012, up from 25 percent a year earlier. If smart phones continue to gain at even this pace, "feature phones" will be largely a memory in another five years[3]. In recent years, many regular mobile devices have screen reading software that allow blind users to be able to use them. [2]. *"Touchscreen devices like the iPhone were once assumed to be inaccessible to blind users, but well-designed, multitouch interfaces leverage the spatial layout of the screen and can even be preferred by blind people "*[13].

The fact that our handheld devices are getting better connectivity options (better Wi-Fi, better Bluetooth), better computing capacity and that their proliferation is increasing each year allows us to develop applications that can communicate with other devices or services allowing the trade of data and localization.

---

[2] http://mobithinking.com/mobile-marketing-tools/latest-mobile-stats/a#smartphone-shipments

[3] http://mashable.com/2012/05/09/smart-phones-spreading-faster/

These capabilities enabled us to design and implement an application that can be introduced in the day to day life of visually impaired people, to provide a sort of a personal assistant which can perform a number of tasks to help out in the normal life of a user. It can be framed as an awareness tool we provide blind users to contribute to the supply of information they can get out of their surroundings. First of all a tool that helps blind users identify and be aware of the presence of other people in their surroundings. A context awareness tool that allows for a wide array of usage, with an embedded notification system which can be used to trade information between users and serve as a data storage, from a social perspective for simple message trading, for work related such as appointments, informational notifications for navigation where the application helps users be more comfortable in new surroundings by knowing where they are and by having access to all sort of helpful information, all this associating people, location or time to these notifications.

A big concern is to be able to provide all this in an accessible way for visually impaired people. We wanted to leave behind, after our work, a set of foundations to what these applications can and should do for visually impaired users. Identifying the important information, the good scenarios where users need the most help and would like to get the most help.

## 1.2 - Awareness Tools

The goal of this dissertation is to create a complete system that can provide information regarding objects, people and locations both indoors and outdoors - a tool to provide information about the surrounding of a person. This tool is to be usable by visually impaired people to allow them to have further knowledge about their surroundings. Concretely, we wanted to achieve a functional and practical application which allows blind users to use it has a tool to navigate new environments, to enable them to be more aware of their surroundings and finally be able to use it to trade notifications in a unique way between users using not only physical locations to tag notifications but also people. The first step we did was to identify through some questionnaires, interview and brainstorming sessions together with our blind participants what should be our main focus when considering what awareness we could provide for a blind person. After we got some results from the data we collected over the interviewing of our blind participants we focused on developing a first prototype to test our findings. We developed a prototype capable of tagging users. This first prototype is able to identify people and associate information with them. This prototype takes into account proximity and is usable both indoors and outdoors. After this the goal is to add the possibility to create information associated with locations and objects. Finally we want to allow blind users to share this information between each other.

The system comprised both outdoor and indoor locations which is something that most applications lack. Most systems focus only on the outside since the indoor ones are much more complex than the simple usage of the GPS coordinates provided for outdoors locations.

Notification systems are usually associated with locations and in some cases they also use time as a parameter, however there is much less work done regarding identifying people specifically and tagging people instead of locations. For the visually impaired there is also the fact that anything that is usually used as a visual note, message board, posters, televisions are not easily perceived, even message boards with braille make them search the board for anything new as they do not know immediately what is being displayed to them.

The applications existing nowadays try and tackle these situations one by one and they do not offer a solution that envisions a complete system that works both outdoors and indoors. Also most of these do not even include a notification system to allow users to trade information between one another.

There is a lot of work done in this area regarding tagging locations mostly on outdoor environments and some work done in indoor environments however there is still a lack in the tagging of individual users instead of places. It is harder to deal with the user position directly because they are usually in movement and so identifying them is a complex problem.

## 1.3 - Contributions

Our main contribution was focused on answering the question if awareness systems can be used to help blind users. And provide information about how to properly implement such a system. During the duration of the project we had the following contributions.

- A State of the art review on awareness tools for blind people aiming to provide context both explicitly and implicitly.

- An in-depth analysis of the limitations and needs of blind people in context-challenging environments. This analysis spanned novel and known spaces, both indoor and outdoor and an overview of their day to day in these environments.

- A distributed mobile prototype able to provide awareness about people and specific locations. This system also allowed users to generate information and save it in the form of notes that could be shared for multiple purposes and in multiple ways.

- A mobile user interface adapted to blind people's abilities designed with a user-centered approach.

- An evaluation methodology for context-aware systems in the wild.

## 1.4 - Publications

The contributions provided in this dissertation were accepted for publication in two peer-reviewed conferences:
- Ivo Rafael, Luís Duarte, Luís Carriço, Tiago Guerreiro, "*Towards Ubiquitous Awareness Tools for Blind People*", In Proceedings of HCI 2013– The 27th International British Computer Society Human Computer Interaction Conference: The Internet of things, London, UK, September 2013.
- Ivo Rafael, Luís Duarte, Luís Carriço, Tiago Guerreiro, "Ferramentas Contextuais para Pessoas Cegas", Actas da Interação 2013 - 5ª Conferencia Nacional sobre Interação, Vila Real, Portugal, Novembro, 2013

## 1.5 - Document Structure

In chapter 2, we describe the related work done in this area and try to cover the most significant projects that involve context awareness and their positive and negative aspects. An analysis of these projects showed that there are lacking mobile systems that provide implicit and explicit context awareness. We make a breakdown of these projects and do a summary of the analysis that we took into account for the next step. This step is described in Chapter 3, where we go into detail about our first steps into figuring out what exactly the blind users need and what they expect regarding context awareness. We explore the limitations by interviewing 13 blind people and probing them with a proof-of-concept prototype able to detect people in the surrounding environment. These studies confirmed the need for awareness tools and brought new scenarios particularly in outdoor scenarios and in augmenting locations with annotated information. Chapter 4 presents the details of the implementation of this system. In particular, we depict the architecture of the system and the non-visual user interface. It is the technological chapter where we try to provide information of how someone could replicate the system anytime. The final chapter, Chapter 5, we discuss the results of our field test made with the application design presented in chapter 4. After a one week and a half field test with the final application we did a series of daily small debriefings followed by a final questionnaires and interviews that we used to collect our results and sum up our findings. We use this chapter to present our findings and what we perceive as guidelines for the development of such tools.

# Chapter 2 - Related Work

This chapter presents a critical overview of previous work pertinent to this thesis. What we explore is the environment context for a person. The awareness of ones surrounding location. People in our presence, identifying our location, perceiving the space around us, objects, what information can be taken from an environment. Low level fragments of context information, for example, that five people are located within a few feet of coordinates (1,10,20) in a building coordinate system, that those coordinates correspond to Conference Room A, and that Conference Room A is currently scheduled for a budget meeting, may be composed to deduce higher-level information, for example, that each of those individuals is currently involved in a budget meeting.

This context information can be associated with systems that take into account location as their primary source, others are focused on objects as means to relay information to the blind user and there are also the systems that focus more on the social aspect, which are more related to the relation between people. The idea behind all these systems is that they want to provide the blind user information about different things in their surrounding using different perspectives.

The next section shows examples of these systems and how they try to help out the blind user be more aware of its surroundings and what context he is inserted into. It will also give an overview about the technologies used in these systems and compare them thru several aspects.

## 2.1 - Awareness Systems

Awareness systems can be broadly defined as those systems that help people construct and maintain awareness of each other's activities, context or status. Context awareness and location awareness are concepts of large importance. The term of location awareness is still gaining momentum with the growth of ubiquitous computing[24]. First defined with networked work positions, it has been extended to mobile phones and other mobile communicable entities. The term covers a common interest in the whereabouts of remote entities, especially people. "*Context awareness is defined complementary to location awareness. Whereas location may serve as a determinant for resident processes, context may be applied more flexibly with mobile computing with any moving entities, especially with bearers of smart communicators*".[4] We will focus on three types of systems based on their focus about the information they try to relay, location systems, object systems and social systems.

### 2.1.1 - Location Related Systems

Some systems focus on location as the main focus of information, most of them are used for navigation or just orientation, which is one of the most important difficulties of blind users in new environments. Having information about locations for the blind can help them feel more comfortable about their surroundings.

#### 2.1.1.1 - Museum Guides

This system developed by Ghiani et al [6], is a multimodal and location aware museum guide, which has been specifically designed for visually impaired people to provide them with flexible orientation support. It brings the benefits of location-aware sensing to blind users by enhancing a mobile guide with reliable orientation support. In this way blind users can access information regarding artworks or scientific specimens in their original locations, even if they cannot directly appreciate them. Such support can assist blind users both when they can touch an item on display and when it is not possible. In fact, if blind users can receive notional and (alternative) descriptive information on the artworks available, they can have a more enjoyable and informative visit. In addition, such support can enhance a blind person's museum visit together with their family and/or friends. Thanks to such support, blind visitors can be more autonomous and socially integrated. This system is designed as an indoor location system; it makes use of RFID technology to accomplish its goals.

---

[4] http://www.answers.com/topic/context-awareness

This system's main advantage is that it works indoors which most do not, and it provides not only orientation information for the blind user but also some regarding particular objects even though it is all based on location.

### 2.1.1.2 – Autonomous Navigation through the City

While navigating through the city, blind people are faced with various obstacles and situations within the urban landscape. Such elements are entirely common for most people, but the situation is different for people with vision impairments. This difference that exists may be a source of fear for blind people when navigating on their own, thus affecting their mental representation of the space traveled. Currently, a verbal description of the environment provided by sighted people is the most commonly used aid for blind people to be able to navigate in open spaces, or that of an experienced blind user in the environment they are in. This largely limits safe and independent navigation by blind people both in their homes, in nearby environments and in outdoor spaces. The software presented in by Sanchez et al [22] *"consists of a simple and low cost software and hardware solution that helps blind users to navigate and carry out their daily outdoor mobility tasks in both familiar and/or unfamiliar environments"*. This tool consists of an audio based application that is integrated into a mobile device, which together with the help of GPS satellites and a GPS device, provides blind users with information in order to be able to orient themselves and navigate through various points of interest in the city.

This system is designed for the outdoor environment and allows the blind user to get information regarding their surroundings and their physical location.

The advantage is that they use audio-based GPS software. This software favors the navigation of blind people in unfamiliar contexts, since it is not a requirement that the user has already been through the route. However the software does not provide enough information to be able to detect certain obstacles in the path, such as curbs, roadwork or storefronts.

### 2.1.1.3 - Integrated Indoor/Outdoor Blind Navigation System

There are many systems that try and help blind people navigate thru new locations, however most of them do not work both indoors and outdoors. This project [19] uses an Original Equipment Manufacturer (OEM) ultrasound positioning system to provide precise indoor location and Differential Global Positioning System (DGPS) for the outdoor location tracking. This system is capable of providing the user with the layout of the indoor facility, and gives a broad picture of what the environment is like. The user may also get distance and navigation information between destinations. Another feature implemented is that of travel safety, as the user walks around the system timely calls out obstacles. The system can also communicate with the user and answer different contextual awareness questions on demand.

The major advantage of this project is the fact that it complements itself by being able to work both outdoors and indoors which is very important. However this system is not so simple and cheap, it does not use a simple device like a smartphone or a tablet, and it requires the user to carry around extra hardware.

## 2.1.2 - Object Related Systems

Another type of systems seeks to provide information about objects in the user's surroundings. This is commonly achieved by the means of object tagging or image-based detection.

### 2.1.2.1 - Wearable Object Detection System for the Blind

Visually impaired people are faced with the need to identify certain items in their day to day life, in order to be able to organize their activities. However this proves to be very difficult in some cases, distinguish similar objects just by touch is not an easy thing to do. RFID uses radio waves to deliver data from a tag, which stores information, to a reader, which can elaborate the information making decisions. This project [4] uses a RFID device designed as a support for the blind in searching some objects he is presented, in particular, it has been develop for searching the medicines in a cabinet at home.

The device is compact and can easily be placed inside of a glove. The device is able to provide to the blind the information stored in the scanned tags and the value of RSSI correlated to a biofeedback signal. This function can be used by the blind to reach an object, to which an RFID tag was applied, by means of an acoustic signal.

The project requires extra devices to be carried by the user which is not ideal, and has issues when multiple RFID tags are detected.

## 2.1.2.2 – Post-IT

Brown[17] describes a framework for creating context-aware applications. A stick-e document is composed of a set of stick-e notes, each resembling a HTML page. Each stick-e note consists of content, and the context in which it will be triggered. Consider a mobile user carrying a PDA equipped with location sensing hardware. The user can place a stick-e note at a physical point of interest. When the user returns to that position in the future, the stick-e note will be triggered, the user being informed of this by the PDA. A stick-e note can therefore act as the electronic equivalent of a Post-it note. Some other examples of context that could trigger a stick-e note are given: the adjacency of a person to other physical objects and when the temperature is below a certain level. The stick-e note approach offers a useful general mechanism for the creation of context-aware applications. Further work appears to be required on system aspects, that is the management of location information and the delivery of contextual information to applications.

## 2.1.2.3 – Color Identification for Blind People

Visually impaired people that need to know the color of some object, or to identify the object by just knowing its color, are able to do so only by asking it to another person, when this one is nearby. This application [5] allows blind people to accomplish such a task independently. This system uses image processing in order to accomplish its goal, image processing requires that the various color models be priory studied. The color information for each pixel can be specified in standard ways, and accepted through these color models. Each model is specified by a coordinate system, generally tridimensional, where each color is represented by a single point. In this work, it was possible to design and implement an application which allows determining and communicating audibly the predominant colors of an image taken by the cell-phone's camera.

## 2.1.2.4 – VizWiz

VizWiz [1] is a talking application developed for smartphones that takes advantage of the web to answer visual questions in nearly real time. It is a project aimed at enabling blind people to recruit remote sighted workers to help them with visual problems. The application is used on their smartphones with a camera, they ask questions and receive answers from recruited sighted users. These tools already existed for the blind but the good thing about VizWiz is that its cost is much lower. Also the fact the application is reliant on real users makes the scope of the questions it can help answer much larger. This project also created an approach to achieve low-latency responses called quikTurkit, the main idea is queuing the workers before they are needed this makes the answers come fast enabling a proper usage of the tool.

## 2.1.3 Social Related Systems

These systems tend to focus on the social aspect of sharing and connecting users in order to relay the information regarding their surroundings. These systems can recur to crowd sourcing in order to accomplish their goals. The good thing about these systems is that it provides the social component that blind users could use, help them communicate and have social connections much more simply. Also these systems are very lacking in the visually impaired department, most of them have no concern for accessibility while others focus on other impairments like mobility.

### 2.1.3.1 - Accessible Contextual Information for Urban Orientation

A number of systems have been developed to provide location-specific information about predefined landmarks for mobile users, these systems rely on static content associated with each location, and thus could not adapt to changing or emerging user needs. To address this, a handful of systems have explored the viability of allowing users to generate arbitrary annotations for physical locations. This is where this project [23] comes in, it allows for visually impaired people to generate orientation notes and share them between one another. This project takes the orientation onto another level by adding a socially maintained online database containing information about POIs. This system however is only developed for outdoor environments which makes it lacking. An urban orientation and contextual system such as this offers relevant, dynamic, and up-to-date information, the combination of which may not otherwise be accessible.

### 2.1.3.2 – OurWay

Accessibility maps are valuable tools for people with mobility problems navigating in the urban landscape, OurWay [9] is a project that lets users create and augment geospatial data. This user generated content provides a basis for computing satisfactory routes, from one location to another, matching the user's preferences and needs. This helps reduce the cost of such systems which are usually hard to maintain and to keep updated. By adding the social component of being able to create content and share it the big question that arises with this system is the trust needed in the people creating the content.

This is a tool developed for people in wheelchairs or with a physical impairment that makes it impossible for them to use certain paths. This type of implementation has not been properly explored for the visually impaired people.

### 2.1.3.3 – PeopleTones

Having a tool that can report situations that are of interest to the user is a very useful thing, PeopleTones [14] is a buddy proximity system. This application can be used to help in a variety of scenarios like arranging meetings. The project tries to focus on a proximity detection algorithm, ways to reduce the noise and interference and finding new ways to share these "good to know" things in the most unobtrusive way but still effective. The user study revealed that peripheral cues are an effective, unobtrusive mechanism for notifying people of such inferences. Although haptics have often been suggested as a promising ambient delivery mechanism, sound was the preferred medium, possibly because of its higher fidelity. This is another area that is not explored for blind users, the possibility of detecting users in their proximity. Of course the method of output is one of the most important things when thinking of an application of this kind for blind users.

## 2.2 – Discussion of Awareness Systems

Most of these projects try to give information regarding one component while some manage to provide support for two types of information. Some of the components have not even been properly explored focusing on the visually impaired, especially the ones that concern the creation and sharing of information.

This is an overview of the projects, what they focus in and what they lack in regards to type of information/identification.

Also some of them never considered the ubiquitous scenarios and only work either indoors or outdoors which is not optimal.

| Project | Object | Location | Sharing | Indoor | Outdoor | For the Blind |
|---|---|---|---|---|---|---|
| Museum Guides | 🟢 | 🟢 | 🔴 | 🟢 | 🔴 | 🟢 |
| Autonomous navigation | 🔴 | 🟢 | 🔴 | 🔴 | 🟢 | 🟢 |
| Indoor/Outdoor Blind Navigation | 🔴 | 🟢 | 🔴 | 🟢 | 🟢 | 🟢 |
| Wearable Object Detection | 🟢 | 🔴 | 🔴 | 🟢 | 🟢 | 🟢 |
| Post-it | 🟢 | 🔴 | 🟢 | 🟢 | 🟢 | 🔴 |
| Colours Identification | 🟢 | 🔴 | 🔴 | 🟢 | 🟢 | 🟢 |
| VizWiz | 🟢 | 🔴 | 🔴 | 🟢 | 🟢 | 🟢 |
| Accessible Contextual Information | 🔴 | 🟢 | 🟢 | 🔴 | 🟢 | 🟢 |
| OurWay | 🔴 | 🟢 | 🟢 | 🟢 | 🟢 | 🔴 |
| PeopleTones | 🔴 | 🔴 | 🟢 | 🟢 | 🟢 | 🔴 |

**Table 1 Comparison of awareness systems**

Table 1 shows a comparative evaluation of the aforementioned systems. Out of all the projects we present only half of them provided some support for object awareness and by contrast the other half presented location awareness. Only one crossed those two features and has object and location awareness. As far as sharing the information that the tool provides again only half of them have this capability. There were some projects that focused only on indoor or outdoor environments although 6 out of 4 did support at least some functionality for both type of environments. Out of all the projects mentioned not all of them were developed with accessibility taken into account, 3 out 7 do not are not blind user accessible.

Most projects are too specific when trying to tackle the problem and so never look at the overall picture as an example you have Museum Guides or Colors Identification, which only focus on specific problems and thus are very limited. Then we have projects that have a wider approach like VizWiz that allow for identification at multiple levels but forget about the other components like being able to create and share information or also take into account location.

## 2.3 – Supporting Technologies

All these applications have a similar technology base that it is used for location, object or people detection and identification. In this section we present and compare the various available technologies and their advantages and disadvantages.

### 2.3.1. - Global Positioning System (GPS)

GPS was conceived as a navigation system. By knowing the position of the satellites and measuring the distance between its antenna and four or more other satellites, a single GPS transmitter can compute its three dimensional speed, position and direction of travel [18].

The system is not perfect and there are errors when calculating the position, the actual position will be in a radius measuring from 20 to 100 meters. Fortunately there are ways to reduce the margin of these errors.

GPS system are commonly used for all sorts of outdoor location systems, the most common are for personal use which can be the base for all sort of applications with different objectives. Recent works done with GPS systems try to go a step further and find ways to deal with or improve the error and the uncertainty that the system inherently has [3]. GPS can be used for Geotagging or Georeferecing in the outdoors, clock synchronization, navigation, Geocaching, etc.

*Advantages*:
- GPS works in all weather
- Relatively low costs (compared to other navigation systems)
- Large coverage around the planet
- Accuracy has a pretty good value for its cost
- Relatively easy to integrate into other technologies

- The system is maintained regularly by the US government (as compared to other navigations systems by other countries e.g. GLONASS)

*Disadvantages*:
- GPS satellite signals are weak so it doesn't work as well indoors, underwater, under trees, etc.
- The highest accuracy requires line-of-sight from the receiver to the satellite, this is why GPS doesn't work very well in an urban environment
- The United States Department of Defense can, at any given time, deny users use of the system (i.e. they degrade/shut down the satellites)

## 2.3.2 - Bluetooth

Bluetooth is a wireless technology through which a user can transfer data between two devices having required proximity. It has now become one of the handiest developments in the wireless technology ground which is now being used for multiple purposes and in a good range of devices for e.g.: cell phones, laptops, headset devices, video game consoles, printers, tablets, music players and HD TV's. There is also an upcoming innovation to the Bluetooth Protocol coming out which is the Bluetooth 4.0 which provides massive improvement to the power consumption.

*Advantages*:
- The processing power and battery power that it requires in order to operate is very low.
- It's very simple to use, anyone can setup a connection and sync two devices with ease.
- The chances of network interference are very low.
- Bluetooth functions at less than 100 meters but it doesn't require a line of vision and is cable free.

*Disadvantages*:
- Albeit the transfer speeds being around 1Mbps it's much slower than other similar technologies like Infrared or WLAN.
- Security is good even though it's not the best when compared to infrared.

- The battery usage during a transfer is negligible but leaving the device switched on can drain the battery life considerably.

### 2.3.3 - RFID

RFID stands for Radio-Frequency identification. The acronym refers to small electronic devices that consist of a small chip and an antenna. The chip typically is capable of carrying 2,000 bytes of data or less. The RFID device serves the same purpose as a bar code or a magnetic strip on the back of a credit card or ATM card; it provides a unique identifier for that object. And, just as a bar code or magnetic strip must be scanned to get the information, the RFID device must be scanned to retrieve the identifying information.

There are two distinct type of RFIDs, the active and the passive.

|  | ACTIVE RFID | PASSIVE RFID |
| --- | --- | --- |
| Power | Battery operated | No internal power |
| Required Signal Strength | Low | High |
| Communication Range | Long range (100m+) | Short range (3m) |
| Range Data Storage | Large read/write data (128kb) | Small read/write data (128b) |
| Per Tag Cost | Generally, $15 to $100 | Generally, $0.15 to $5.00 |
| Tag Size | Varies depending on application | "Sticker" to credit card size |
| Fixed Infrastructure Costs | Lower – cheaper interrogators | Higher – fixed readers |
| Per Asset Variable Costs | Higher – see tag cost | Lower – see tag cost |
| Best Area of Use | High volume assets moving within designated areas ("4 walls") in random and dynamic systems | High volume assets moving through fixed choke points in definable, uniform systems |

| Industries/Applications | Auto dealerships, Auto Manufacturing, Hospitals – asset tracking, Construction, Mining, Laboratories, Remote monitoring, IT asset management | Supply chain, High volume manufacturing, Libraries/book stores, Pharmaceuticals, Passports, Electronic tolls, Item level tracking |
| --- | --- | --- |

Table 2 Active vs Passive RFID - http://www.atlasrfid.com/auto-id-education/active-vs-passive-rfid

*Advantages*:

- Doesn't require line of vision to work
- Signal is not blocked by common materials
- Can have a reach of several meters
- Several tags can be read at once by a single reader
- RFID tags are very simple to install/inject inside the body of animals, thus helping to keep a track on them. This is useful in animal husbandry and on poultry farms. The installed RFID tags give information about the age, vaccinations and health of the animals.

*Disadvantages:*

- Though it is very beneficial, it quite is expensive to install. Small and medium scale enterprises find it costly to use it in their firms and offices.
- It is difficult for an RFID reader to read the information in case of RFID tags installed in liquids and metal products. The problem is that the liquid and metal surfaces tend to reflect the radio waves, which makes the tags unreadable. The tags have to be placed in various alignments and angles for taking proper reading.
- Transmission rate is not very high

### 2.3.4 - Ultrasound

Ultrasound is acoustic (sound) energy in the form of waves having a frequency above the human hearing range. The highest frequency that the human ear can detect is approximately 20 thousand cycles per second (20,000 Hz ). This is where the sonic range ends, and where the ultrasonic range begins. Ultrasound is used in electronic, navigational, industrial, and security applications. It is also used in medicine to view internal organs of the body.

*Advantages:*

- Efficient when used for obstacle detection

*Disadvantages:*

- Very prone to interferences

## 2.3.5 – Infrared

Infrared detectors are a modern technology used to pick up an area of the light spectrum that the eyes are not capable of seeing. Also known as "thermography", using infrared detectors has a variety of uses in today's society, including, construction, public service and science.

*Advantages:*

- Their ability to be applied to a large area. Detectors can be used in much the same way that eyes can to survey an area and pick up the infrared section of the light spectrum.
- Operating in real time, infrared detectors pick up movement making them useful in a variety of circumstances

*Disadvantages:*

- Because infrared detectors detect infrared images based on the temperature variants of objects they cannot detect differences in objects that have a very similar temperature range. This leads to inaccuracy in many circumstances.
- Infrared detectors are extremely expensive, which limits their use in many sectors.
- They usually required vision between the devices in order to work.

## 2.3.6 - WLAN

A wireless local area network (WLAN) is a local area network (LAN) that doesn't rely on wired Ethernet connections. A WLAN can be either an extension to a current wired network or an alternative to it. WLANs have data transfer speeds ranging from 1 to 54Mbps, with some manufacturers offering proprietary 108Mbps solutions. The 802.11n standard can reach 300 to 600Mbps. Because the wireless signal is broadcast so everybody nearby can share it, several security precautions are necessary to ensure only authorized users can access your WLAN. A WLAN signal can be broadcast to cover an area ranging in size from a small office to a large campus. Most commonly, a WLAN access point provides access within a radius of 65 to 300 feet.

*Advantages:*
- WLANs allow mobility and availability, as you can take a device anywhere in the house without plugging in.
- It is cheaper and easier to add new devices to a network, as there is no need for any more wires or cables.
- Transmission rate is quite fast

*Disadvantages:*
- The bandwidth is much lower than standard cable bandwidths, and WLAN is also less efficient and reliable.
- There is less security from malicious attacks, and interference can be caused by other radio signals, which can result in loss of signal
- Installation and maintenance cost might be too high depending on the coverage of the WLAN

## 2.4 Discussion about technologies

The new mobile computing applications must take advantage of the fact that they are able to provide us with the physical location of things and give us that information [8]. Researchers are working to meet these and similar needs by developing systems and technologies that automatically locate people, equipment, and other tangibles.

Many solutions have been developed over the years, each tries to solve a different problem or support different applications, they turn out different from one another they vary in many parameters, such as the physical phenomena used for location determination, the form factor of the sensing apparatus, power requirements, infrastructure versus portable elements, and resolution in time and space.

These location systems can be characterized by a set of properties that we use to compare the systems presented above.

➢ <u>Speed</u>

These systems have different speeds when we think about transmitting data. Depending on the amount we want to use this might be an issue. Or at least something that will differentiate the usage of one system over the other.

➢ Accuracy and precision

These two properties go hand in hand, one relates to how often is the location correct and the other is by how much distance can we be certain of that location. Cheaper GPS receivers can give us a precision of +/-10 meters and an accuracy of 95% while the more expensive ones can reach up to +/- 3 meters and 99% accuracy.

➢ Scale

To assess the scale of a location-sensing system, we consider its coverage area per unit of infrastructure and the number of objects the system can locate per unit of infrastructure per time interval.

➢ Cost

There are several costs involved in a location system. The time cost takes into account the installation process and the system administration needs. Space costs for the amount of infra-structure. Capital cost for the price of all the needed hardware.

➢ Limitations

Some systems will not function in certain environments. One difficulty with GPS is that receivers usually cannot detect the satellites' transmissions indoors. This limitation has implications for the kind of applications we can build using GPS.

There are two big types of navigation systems, the ones that focus on the indoor environment and the ones that focus on the outdoor environment. The most common and easy to work with is the outdoor environment, the GPS (Global Position System) is a very practical and established system that allows for easy location tracking thru coordinates on the outdoor environment however the GPS system does not work properly indoor which makes the systems and techniques used in this environment a much more open debate.

If we take into account the visually impaired users these systems need to be carefully thought due to the fact that these users are prone to be put in dangerous situations if the system does not perform well, even though they are self-sufficient and usually trust their walking cane and remaining senses above all else.

| Technology | Speed | Accuracy and Precision | Scale | Cost | Outdoors/Indoor | Hardware Requirements (Smartphones capabilites) | Line of Sight |
|---|---|---|---|---|---|---|---|
| GPS | ~50bps | 10 meters | Good | Good | Outdoor | Almost all | No |
| Bluetooth | ~1Mbps | 100 meters | Average | Average | Both | Almost all | No |
| RFID | ~100-8000bps | 10cm- 200 meters | Average | Average | Both | Some | No |
| Ultrasound | --- | Centimeters to meters | Average-Bad | Bad | Both* | Few | No |
| Infrared | ~115kbps | Centimeters | Bad | Bad | Both | Some | Yes |
| WLAN | ~54Mbps | 40-90 meters | Average | Average | Both | Almost all | No |

**Table 3 Table of comparison of technologies**   - Good   - Average   - Bad

# Chapter 3 - Approach: Ubiquitous Awareness Tools for Blind People

Visually impaired people find themselves constantly needing more information regarding their surroundings. They are confronted with several scenarios where a simple tool that would help them get some more information would be extremely helpful.

We decided to proceed to find out what was lacking in a blind user day to day life, what their difficulties were or how they handled some normal situations that required some extra care considering their visual problem.

This meant making an approach focused on the user. We decided to approach our target population and focus our research around them. We produced a set of questionnaires and interviews as well as some brainstorming sessions for us to understand their difficulties and to help us focus our work into the issues that mattered the most to them.

After researching the state of mobile awareness for blind people we managed to identify some areas of interest that we wanted to explore to try and provide tools that could help manage some scenarios that, we had to make sure though that our findings were justified with results from data collected from the real users we wanted to target.

The lack of information blind users get about their surroundings can be impairing to them in some situations, that information can be of different types, it can be about their location where they are, what is the place they find themselves in, it can be about people where are they, how many are there, who are they or even objects what is it, where is it, shapes and sizes. Our first approach focused first on trying to confirm some of these issues and try and provide a platform to generate new scenarios or new situations where a blind user felt he lacked the required information to deal with.

For this we decided to first to an interview approach followed by a development of a first prototype to allow us to confirm that our concerns and the problems we found were valid.

## 3.1 - Evaluation Methodology

For the first steps into our project field work we decided on a three phased approach. First we started with a pre-questionnaire in which we aimed to understand the limitation of the notion of awareness or context awareness and environment awareness for blind users. We also aimed to describe our test group and its characteristics while at the same time providing them with some scenarios, problems or possibilities they would find useful or interesting that we could explore and try to work upon. Finally we also wanted to try and understand their technological usage on a normal day. These questionnaires had the objective of providing us with some immediate confirmation that there is a problem and that in fact there is something that can be done to alleviate this problem. We needed to access with the blind users that the information we pretended to give them was indeed useful and needed and not just a tool to be cast aside with features they were not interested in or didn't need. These questionnaires were accompanied with some brainstorming sessions where users where free to express any opinions or thoughts they had on the issue presented to them, this part of the approach intended to provide us with more data concerning what could be a good scenario to explore in our study. The experience some of our users had with technology allowed us to have some information as well about some key design features blind users are used to in whatever applications they use.

Based on our findings, and taking in consideration the state of research, we focused our attention on awareness about people and information about them. We developed a first prototype and our goal with this prototype was to elicit new scenarios and, together with the users, find new settings where mainstream technology could support them by adding layers of information to the knowledge they have about their surroundings.

The first area we focused in was providing blind users with information about people in their surroundings and an accessible notification system they could use to create notes or reminders for themselves.

The project designated UAT - Ubiquitous Awareness Tools for the blind was developed as a mobile application for an Android Smartphones with Bluetooth capabilities. The application featured a proximity detection system for users which can warn about people in the surrounding environment and a notification system that allows users to create notes and be warned at their convenience about them. Using the unique MAC address of each device's Bluetooth the application can associate that ID to any contact that is present in the user's phone. When using the application, it periodically searches for Bluetooth devices in its vicinity, when one is detected it goes through its database to search for the ID detected to figure out if the device detected belongs to someone known to the user or not. The drawback of this system is that it requires users to have Bluetooth connected and in discoverable mode in order to be able to be detected.

The notification system that comes along with the application is also very simple to use, it allows for the recording of audio notes or the writing of text notes. These notes are created always associated with one contact already added on the application's database. Whenever that contact/device is detected the application goes through the notes it has in order to check if there is any note present and if so warn the user about it.

This first prototype we tested has some limitations concerning the features implemented. First we only allowed the application to warn about devices in the proximity of the user if they are already added as known devices, the notification system was only a personal system has it had no way of sharing these notes created and thus it only served as personal notes. We went to a foundation for the blind where there are a lot of users we could use for our probe and also movement from these users to properly test out the features of the application. It was done mostly in the indoor environment, though sometimes it involved the outdoors as well. This field test intended to evaluate how the application fairs in terms of accessibility, behavior, and features and of course provide new insight as to the possibilities of its usage as well as confirm its usefulness.

Finally after the probe we gathered the users who tested out the application and did small questionnaires regarding accessibility and usage of the application and did brainstorming session with these users in order to come up with new scenarios or new features that we could implement that they find it could benefit them. This was also very important for the next stage of the project not only we needed to confirm our work done so far we needed to guide ourselves into our next objective.

## 3.2 - Participants

We recruited 19 participants for the pre-questionnaires and out of those 15 had cellphones with the needed specifications for the field test which meant we had 15 people participating in the field test. Of these 15 most were only providing their personal Bluetooth ID and were only being used as people that could be detected around the location where the test was being done. We had three actual participants which had our devices with the application installed and they were the ones who had the full experience of the prototype.

## 3.3 - Material

All interviews/questionnaires and the brainstorming session were recorded for a better capture of the information.
As for the field test we used three Android devices, two Galaxy Ace smartphones with Android 2.3.3 and Android 2.3.5 and a Galaxy Mini with Android 2.3.4. All had installed our application and were running it thru the entire duration of the test. For the devices used for being detected we used the users and the participants personal devices all of them Nokia's cellphones with Bluetooth capabilities.

## 3.4 - Procedure

We had two separate days for the first pre-questionnaires where we split almost evenly the number of people per each day. After the pre-questionnaires were finished we proceeded to a full day for the field test which started with a brief explanation and introducing the application to the participants of the test. After about an hour of explaining the features and functionality of the application the participants were left with the devices for the remainder of the day to freely explore the capabilities of the application. During this time they always had if necessary the possibility of reaching out to us if any complication arose so that we could fix it. After the day ended we recovered the devices that stored all the activity they had during that day, from what the users did inside the application as well as what did the application process during all the time it was working, devices found, people found etc.

After recovering the devices we briefly analyzed the data we had to have some guidelines to orient our questionnaires/brainstorming on the final day of the tests. The final day we did a post questionnaires focused on the use and the performance of the application followed by a brainstorming session orientated with the help of some of the data already picked up in the previous day. This session allowed us to get an opinion from the users who tested out the application as well as insight to what they believe the features of such an application should be.

## 3.5 - Results

We will now present the results of our testing and what conclusions we have drawn from each step of the process, from the pre-questionnaires to the final brainstorming session.

### 3.6.1 - Pre-Questionnaires

These questionnaires were oriented for us to try and understand the way our users interact within the scenarios we envision s and deepen our knowledge of their situation and how they adjust to special circumstances. We focus on understanding how they use the technology and how they behave themselves on a normal basis when facing these scenarios. We tried to understand the concept of context for the surroundings of a person, what they find relevant, what they rely on to navigate themselves or have interaction with the world.

### 3.6.2 - User Profiles

The users we interviewed were all visually impaired users; more than 80% of them had full blindness while the rest had light perception at most. The age distribution was pretty diverse ranging from the age of 18 to more than 60 years . Their degree of education varied from the 4$^{th}$ grade up to College Graduates. Overall, the pool of interviewed users was very rich in diversity.

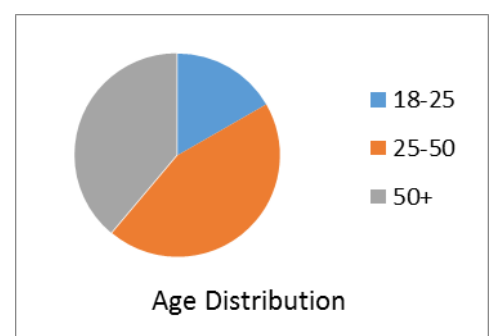

**Figure 1- Pie chart of Age distribution of participants**

We asked questions about the normal usage of technology by these users and what were their usual habits in their day to day life, what sort of devices they usually use, what sort of features from some devices to they use that help them, if they use or had at least tried touch screens at all along with some questions regarding their usage of Bluetooth.

Out of the regular devices usually used the computer, watches and cellphones had a clear presence while all the interviewed had a cellphone and used it only around 66% of them used the computer and/or a watch in a daily basis. Devices like the Braille Machine or others were very rarely used some mentioned the use of a recorder or a mobile drive. None of the enquired used GPS devices or cameras.

| Use of Computer | Cellphone | Watch | GPS | Camera | Braille Machine | Other |
|---|---|---|---|---|---|---|
| YES | YES | YES | | | YES | NO |
| YES | YES | NO | | | NO | NO |
| YES | YES | YES | | | NO | NO |
| YES | YES | NO | | | NO | NO |
| YES | YES | NO | | | NO | NO |
| NO | YES | NO | | | NO | NO |
| NO | YES | YES | | | NO | NO |
| NO | YES | YES | | | NO | NO |
| NO | YES | YES | | | YES | NO |
| YES | YES | YES | | | NO | YES |
| YES | YES | YES | | | NO | YES |
| YES | YES | YES | | | NO | YES |
| NO | YES | NO | | | NO | NO |
| YES | YES | YES | | | YES | NO |
| YES | YES | YES | | | NO | NO |
| YES | YES | YES | | | YES | NO |
| YES | YES | NO | | | NO | NO |
| YES | YES | YES | | | NO | NO |
| 12 | 18 | 12 | 0 | 0 | 4 | 3 |

Table 4. User pool technological profiles

We also enquired about the usage of the touch screens if they had any contact with this type of interaction and how if so how did they feel about it. How hard it was if it felt comfortable or not, easy to learn or not.

Results showed that 50% of them had indeed tried out this technology most of them through tests, meaning they do not possess this type of cellphones only tried it briefly. Out of that half that did had contact with the technology again almost half of them felt this technology hard to use or presented some sort of difficulty.

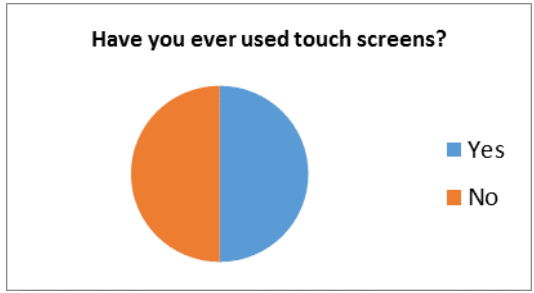
Figure 2 - Touchscreen usage

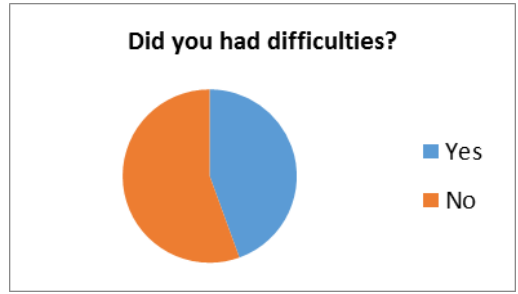
Figure 3 - Difficulties using touchscreen

Considering the cellphone is what we were developing our application for at the moment we also tried to enquire about what features of the cellphone the users take advantage of the most.

| Calls | Messages | Clock | Alarm | Agenda | Contacts | Social Networks | File Sharing | Music |
|---|---|---|---|---|---|---|---|---|
| X | X | X | X |  | X |  | X |  |
| X | X |  |  |  | X |  |  |  |
| X | X | X |  |  | X |  |  |  |
| X | X | X |  |  | X |  |  | X |
| X | X | X |  |  | X |  | X | X |
| X | X | X |  |  | X |  |  |  |
| X | X |  | X |  | X |  |  |  |
| X |  |  |  |  | X |  |  |  |
| X | X | X |  |  | X |  |  |  |
| X | X |  |  |  | X |  |  |  |
| X | X | X | X |  | X |  |  |  |
| X |  | X |  |  | X |  |  |  |
| X | X |  | X |  | X | X |  |  |
| X | X | X |  |  | X |  |  |  |
| X | X |  |  | X | X |  |  |  |
| X |  | X |  | X | X |  |  |  |
| X |  |  |  |  | X |  |  |  |
| X | X |  | X |  | X |  |  |  |
| 18 | 14 | 10 | 4 | 2 | 18 | 1 | 2 | 2 |

Table 5 - Usage of features from the cellphone

Besides using the cellphones for their obvious purpose of calling people on their contact list the enquired users reveal to also use some of the other features. Depending on their need and how easy they are to use. Most of them are able to trade messages through the cellphone even though some still find it quite troublesome. Some take advantage of the clock in their cellphone to know the time although most of them do not use any kind of alarm or agenda to schedule any type of notification for themselves, from wake up calls to doctors' appointments or simple reminders most of them do not use the cellphone for this purpose. Some of the more experienced users also used it for file sharing and listening to music.

### 3.6.3 -Awareness

Part of our questionnaire revolved around awareness questions we tried to find out aspects regarding awareness for blind people in new and old environments, how they guide themselves, the difficulties they face that are common or the things that make it easier on them to navigate on their own, basically anything that can help us realize what is indeed good information that they need and want about their surroundings and what they consider it is unnecessary or just too much.

Most of the interview revolved around awareness questions particularly pertaining orientation in new and old environments, difficulties faced with social interactions and workarounds used in their daily lives. We asked about what sources of discomfort they are usually faced with and which of them are more disruptive.

The main cause of discomfort was the lack of knowledge about the people surrounding them, both whom and how many. Interestingly enough this aspect was majorly stressed by late blinders.

In general, all participants stated that one of their first actions when reaching the formation centre was to ask the receptionist about the presence of their closest friends and if he knew about their whereabouts.

Early-blind people, on the other hand, reported to be comfortable with the unawareness about people nearby. On the other hand, despite their experience, they reported to be uncomfortable with the lack of knowledge about their location and the spatial relationship to objects and places around them, in places that they don't know well or that had *things* moved.

Some users reported that they could infer useful information from what they hear but that they felt lost when there was too much noise, disrupting their sense of navigation and perception. As expected, users reported that new environments were more demanding in regards to attention needing for a search for reference points. One participant stated that he commonly finds safe returning points where he knows he can back up to if he gets lost. The amount of free space in these new environments is also a factor; they claim big open spaces are harder to navigate as well as cluttered places, with too many objects, which damage the acoustics of the room and thus damage echolocation.

When questioned about the impact in changes in the environment, most participants pinpointed misplacing or moving objects out of place.

Pertaining note taking and prompts, the most tech-savvy participants stated to resort to the mobile device to store notes or reminders.

All participants stated to be eager to have deeper knowledge about their surroundings although some stressed that they would desire the tool to be inconspicuous and subtle.

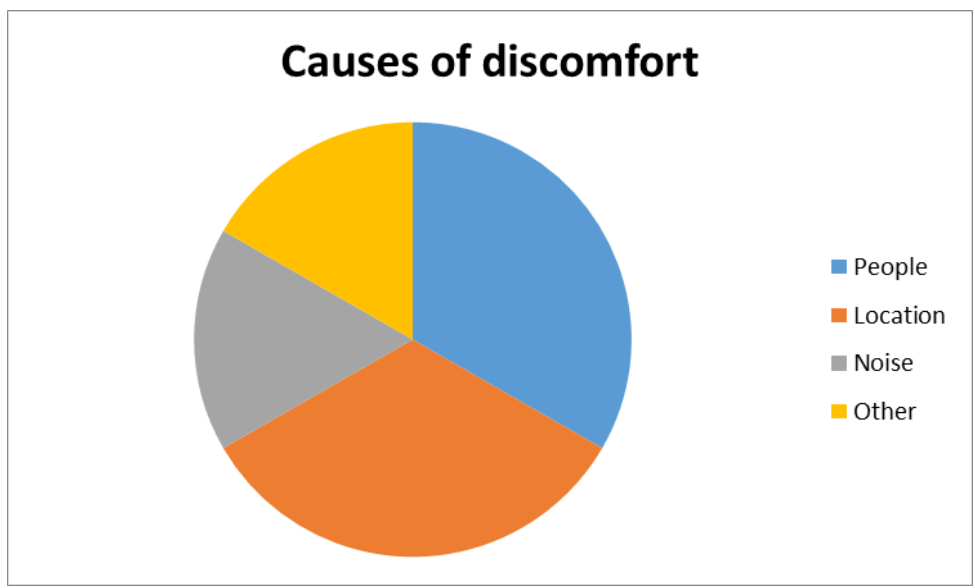

Figure 4- Causes of discomfort on a environment

Another question was made in regards to the differences between old and new environments, how to they navigate differently in them, what is the biggest difference for them when they are in these different types of environment. It was mostly pointed out, besides the fact that new environments are harder to feel comfortable, that new environments require them to focus more and be more alert, they need to search for reference points to be able to navigate in these places, one even pointed out the fact that it is important to have somewhere they can return to if they get lost, that brings them more safety in order to explore if they know they can return to some origin point whenever they feel the need to. The amount of space in these new environments is also a factor, they claim big open spaces are harder to navigate as well as very clustered places with too many objects which usually means bad acoustics for the room which is the most important aspect to them. Worst case scenarios some say some spaces need special help from someone in order to introduce them to the place so they can get their basic references down for navigation.

We questioned them how aware are they about people in their surrounding if they can easily perceive when someone leaves the area or enters it, a room or another type of space. Opinions are divided, about 20% said outright yes and about 25% said outright no they can and cannot perceive people entering and exiting their space. The remainder pointed out that yes they can perceive with several degrees of difficulties. These difficulties or conditions vary between, yes if they can actually hear the person leaving, meaning if someone sneaks out or there is a burst of noise this goes unnoticeable, another pointed out cause is that known people are easier to identify due to sound or movement patterns they are used to but if it is an unknown person they have a harder time perceiving it, finally some said that yes they can always tell when someone leaves or enters but sometimes they cannot identify who that person is.

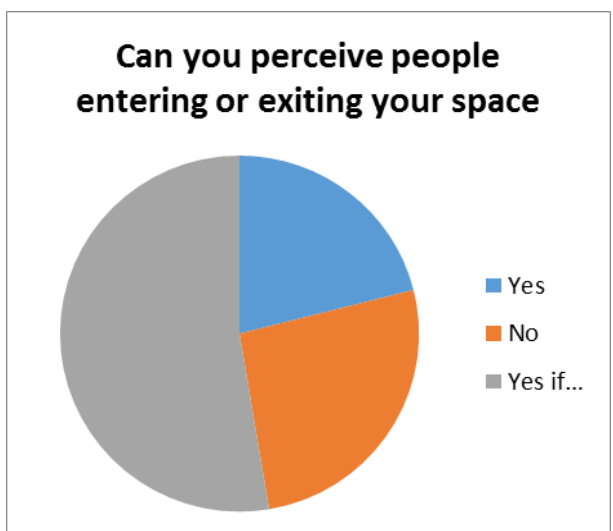

**Figure 5 - Perception of people movements**

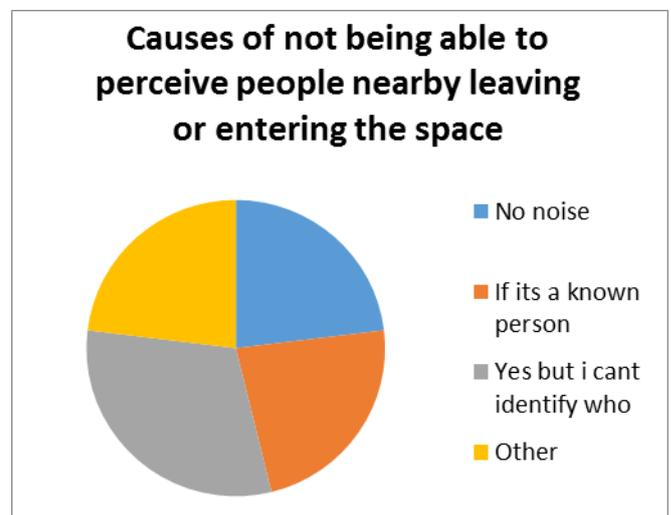

**Figure 6 - Causes of lack of perception on people**

We questioned our participants about what particular changes in the environment around them they have more difficulty in perceiving, the answers to this were a bit ambiguous and only a few had clear issues with something in particular most could not pin point new or interesting things. The most common mentioned was misplacing or moving objects out of place. They either did not mention anything in particular or simply pointed out impairments that could prevent them from perceiving things like noise or ample spaces.

Another question that did not provide too many good answers was when they were asked to identify a particular reason on why some environments are harder to navigate then others or why some are easier to learn or get used to. Even though the answers were also a bit ambiguous some were able to point out that reference points are very important in being able to get used to a place and also the space characteristics if it's an open space or a smaller space, if it is clustered with objects or has good room to move.

Finally we also enquired our participants if they had any tool they utilize to have reminders or notes to themselves or others about appointments or tasks they wanted to remember to do. Over 66% used only a mental note in order to keep track of these things, there was a small number of people that had request the help of others in order to keep track of these affairs while others usually the most proficient were able to use the cellphone to record these events and be notified by the cellphone.

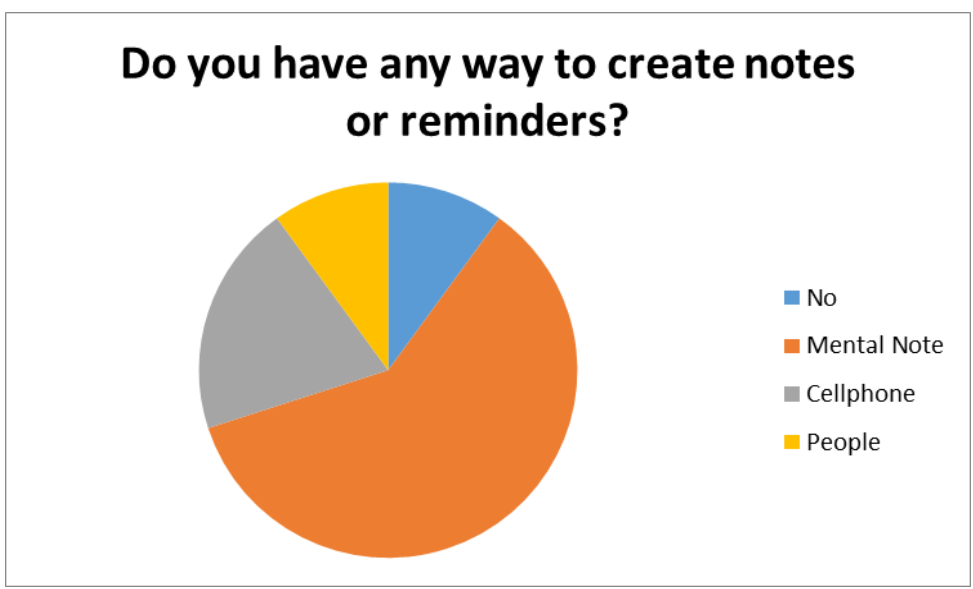

**Figure 7 - What they use as reminders/notes**

## 3.7 –Proof of concept prototype

The prototype was evaluated with three users and it consisted of a full day of use, with the initial hour of users getting used to the application and the devices, with our help to give them the first steps into not only our applications but smartphones and touchscreens in general as it was something most of them had never used extensively. We had a small session explaining how they could use the application and its features followed by a 30 minutes of freely exploring and raising questions about the application and its features and interactions.

After they were comfortable to be left alone with the devices they had the rest of the day for free, unaccompanied but always backed up by onsite support if something went wrong.

The prototype was developed in Portuguese and used the SVOX TTS in order to relay information to the users. The interaction was mainly made with simple gestures and taps with these actions the users could navigate through all the options and features available in the application.

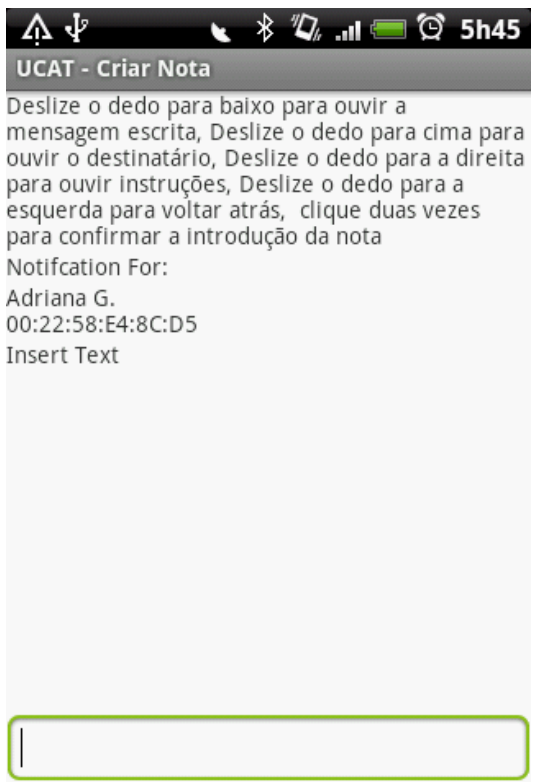
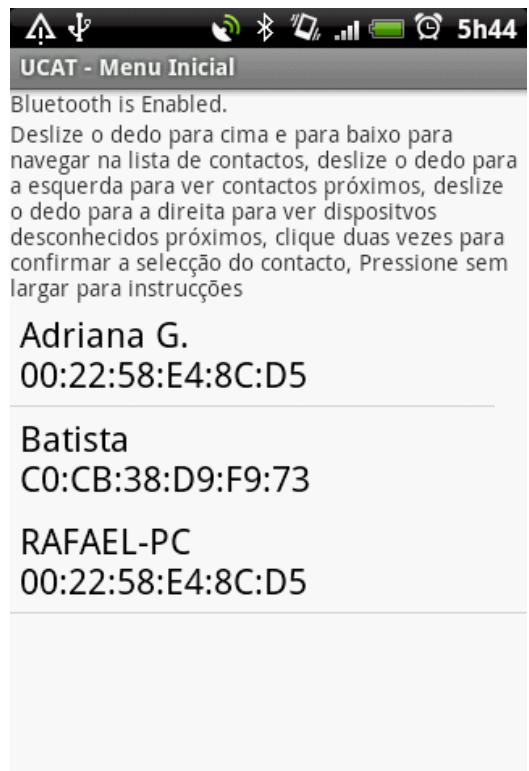

Figure 8 - Note creation

Figure 9 - Initial screen

On the initial screen (Fig. 2) the users had access to their contact list and they could easily access all the options already implemented in the first prototype like creating notes for these contacts or accessing existing notes. The users had the possibility of visualizing all the notifications that the application generated for them from the proximity of people or notes.

To allow for the detection of people nearby, the system uses Bluetooth as a discovery mechanism. Using the unique MAC address of each device, the application can associate that identifier to any contact that is present in the user's phone. When using the application, it periodically searches for devices in its vicinity; when one device is detected it goes through the application database to search for its identifier to figure out if it belongs to someone known to the user or not. The user can navigate through the list of recognized and unrecognized devices, associate identifiers to contacts in the phone, and see when someone was nearby. Further, the application features a notification system; the user is notified when a new device is recognized. The device filters fast on and off identifications to reduce notifications but no more filtering settings were included.

Notifications are offered via vibration patterns. At this point, personalized patterns have been planned but not implemented in this version.

The obvious drawback of the system relies in the usage of Bluetooth and the need for people to have it in discoverable mode. However, our goal here was to assess the benefits of such an approach. Future solutions can resort to other technologies like Wi-Fi and the usage of a server.

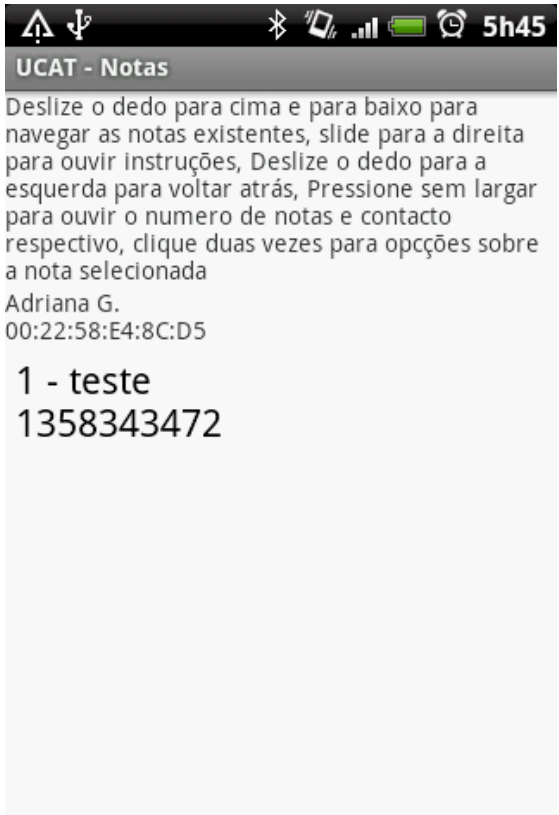

Figure 10 – Notes

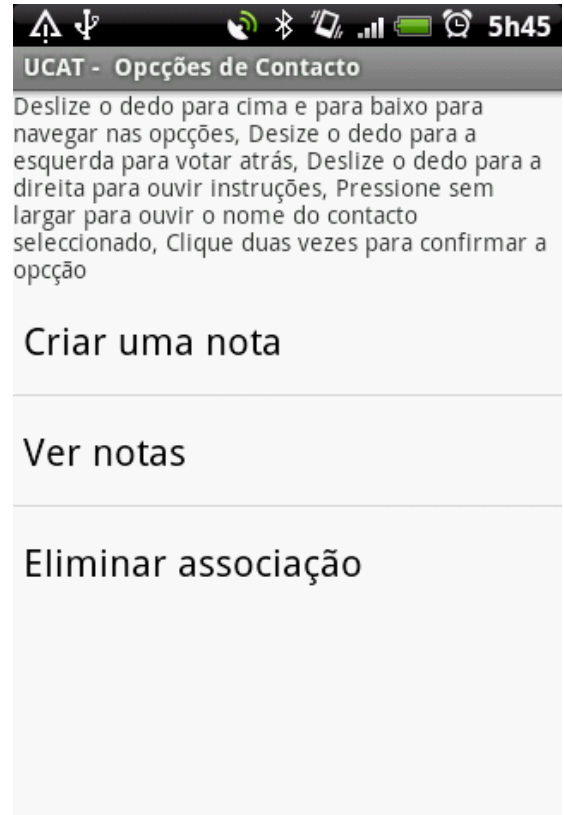

Figure 11 – Contact Menu

To augment the information stored about people nearby, we included the possibility to append notifications to contacts. The system allows recording audio notes or writing text notes. Whenever a contact/device with associated notes is detected the feedback to the user is slightly different (different vibration pattern). The user is then able to receive feedback of people nearby and read notes associated with that contact.

The inclusion of this feature in the prototype was performed to elicit behaviours of augmentation of information to people but also to assess the interests in having notes associated with places and objects.

### 3.7.1 - Usage

The testing was done without too many complications for the first part of the day although as the time went on there were some battery issues due to the devices used already having some batteries not in perfect condition as well as problems maintaining the application on the front of the device. Sometimes the users involuntarily closed the application and they had some issues coming back to the application, this was a problem we had foreseen hence the onsite help to deal with it. We had several approaches to deal with this from locking up some keys to not allow the application to be closed but even with all these measures it was something that happened occasionally.

All three users managed to navigate the application with ease, and it behaved as expected for the most part. Users were detecting the known and unknown devices throughout the all duration of the test and the notifications were being received without problems.

On this stage of the text the users did not experiment much with the notification system and the creation of notes therefore we couldn't gather much data regarding this feature.

### 3.7.2 - User Opinions

The users were satisfied with the fact that the application did provide them with information regarding people that were close to them and that it was able to identify them. They felt compelled to explore the area more in order to try and find people that were in the building.

They were pleased with the overall interaction with the application, the voice was overall good with some minor pronunciation difficulties but most just funny to hear and not at all incapable of being understood. The commands used for navigation didn't take long to get used to even though at the beginning they were not very comfortable or sure on how to perform the swipes or tap movements on the touch screen in order to be recognized.

There was some minor complaint about the lag that sometimes the application in the voice feedback, as they are used to a much faster response from their usual cellphones.

### 3.7.3 - Post-Test Brainstorming

This session provided some insight on what the users thought was good in the application and what was lacking or what features or functionalities could or should be added to provide a more rich application that could provide more information and better performance.

### 3.7.4 – Post-test Interviews

The questionnaires made after the probe revolved around the performance of the application and how the users interacted with it. If it was able to fulfill its objectives, if the feedback provided was sufficient and accurate enough for the users how was the overall usability of the application and some open questions about motivations or issues that prompted some information about the usage of the application.

We asked our three participants if they had more information about their surroundings through the duration of the probe and all of them answered positively, while also confirming to have an easier time identifying when people arrived at a room even though sometimes due to the timing before each search of the application the warning was a bit late and they noticed the person arriving before the warning from the application came. They all found the information useful but also pointed out that sometimes it was too much to the point where they started doubting what exactly was the warning about considering there was too much going on at the same time.

As far as feedback goes they answered positively when asked if the vibration and voice feedback was easily understood and useful. Also the instructions scattered throughout the screens of the application where also easy to understand and important for them to not get lost while using it. Confirmation of actions was also not an issue in almost all situations, some interface glitches caused some confusion in one case. Overall they found the feedback sufficient and good.

As far as the gestures used and interacting with the application no major issues were pointed out. They answered positively when asked if the gestures were intuitive and easy to do. We had a participant try and do a full contact association to a device even though it was not part of the planning and it went without a problem. There were some issues pointed out with notifications, namely the fact that they refreshed if the same person was spotted twice, which meant they did not had an historic of all the notifications but just the last notification, also the fact that they could only remove them one by one was pointed out as a problem. When asked if they had problems navigating the screens or undoing actions they did they all answered they had no issues.

When we moved to the open questions we asked if at any time during the day the notification created a behavior change, all of them said that no. The fact that they were informed of the presence of other people did not prompt them to take any particular action besides interacting with the application. We asked if there was any demoralizing factor during the use of the application and they also answered positively concerning the amount of vibration feedback they were getting. It got to the point where they were constantly being notified about people coming and going which meant they started to doubt the accuracy of the notifications. When asked what type of information they would like to receive about the people near them they mostly answered just more detail about when they were detected, having a history for each contact that they could see.

### 3.7.5 - Group discussion

On this part of the test we had the three participants grouped together and tried to provide them with scenarios or topics to spur the thinking of new or interesting ways that the application could be improved on. We had some suggestions in several areas from the usability or navigation of the application as well as features to implement that could be very useful.

One of the most pointed out factors was the need for more configurations, for example one of them said they would like "a sub list of contacts where one could chose who to be warned about and who to ignore" and even "we might just want it to be on in certain times of the day". Also regarding the notifications they could only navigate them one by one and check or erase them one by one, they pointed out the fact that there should be options to ignore certain notifications or simply quickly dispose of them all and not have to do it one by one.

To some degree they were satisfied with the overall interaction with the application but had a suggestion to perhaps have one or more commands added in order to be able to do more in the same screen without having to navigate so much to accomplish some tasks.

The major suggestion made in regard to features to add to the application focused on the outdoors environment, some mentioned the ability to identify events or obstacles that happened on the street so that others that would follow them on a later time to the same place could be warned about navigation issues or to a more general use be able to leave tips of how to navigate in new environments like knowledge of reference points that they rely on.

Other suggestion involved providing more information about places they go into, one said "It would be nice not having to ask for what exists in a menu when reaching a restaurant", other suggestion involved the stop signs on the street although some of them have a sound system to help out the blind most don't or simply are not the best way to help them one of them suggested a way to relay this information directly to the application so that they knew if the sign was green or not.

Being able to share the notes was evidently pointed out by them, even though they did not get to use them much they provided insight on how they would use them, to warn their families or friends, for example on simple things like reminders to buy something from the local food store if they pass by it.

## 3.8 - Discussion

In this section we try and discuss the results of all the testing of the prototype from the questionnaires to the probe and the post interviews. What where the main points focused as faults what were the good aspects, what needed improvement and what was our focus to implement the final system that we would then take to the test phase once again.

### 3.8.1 - Major Challenges

Concerning the application itself and the use of the components we had some issues to resolve. First the lack of reliability on some Bluetooth devices that tends to behave strangely which caused constant detections which were being interpreted as people leaving the area and returning.

As far as features for the application are concerned, we had several that were suggested and some that we believe will improve its overall performance and usefulness

- Introduction of a personalization system that allows users to customize their contact list and create groups of people whom they want or do not want to take notice of. This resolves issues with privacy and the overload of information that sometimes is unnecessary.

- The ability to share the notifications created by users, this will allow for the users to be able to create notes not only for themselves but also for others. This opens a lot of possibilities for the use of the application notes for more than just personal reminders.
- Several features that include information about the outdoors could be introduced. Some were mentioned by our users like restaurant menus or traffic light information. We have to think about on what exactly we can and cannot do and how to do it. First we want to take the first step to include outdoor locations as a possibility associating the notes which adds to the previous point on how the application note system can be used for more possibilities.
- More information regarding the contacts and the notifications. The first is to allow some more integration with the phone, allow users to see recent messages or phone calls made to the contact they are interacting with. The second is meant to allow the users to receive more information about the application working, history of detections made, being able to do more than just see the notification and no options.
- Introducing indoor locations as points of interest for creating notifications, this means the users can use more than just people to trigger notifications. Using building or office entrances allows for more options about how and when to trigger notifications.

### 3.8.2 – Scenarios

The aforementioned studies enabled us to have a more profound understanding of the awareness requirements of blind people and sharpen our preliminary ideas. Scenarios that arose from our post-test reflections are depicted below

### 3.8.2.1 – Associating notes and reminders with outdoor locations

*John realizes he needs to buy some milk from the supermarket he picks his cellphone and enters a note associated with the supermarket near his house. The next morning John is on his way to work when he walks by the supermarket and receives his notification reminding him to buy some milk.*

This scenario shows how the application allows the user to create a note associated with a location in the outdoor environment using the GPS coordinates, this allows for the possibility of tagging locations with simple self-notifications or even notifications meant for others.

### 3.8.2.2 – Associating notes and reminders with contacts

*John wants to talk to Jules about their plans for the weekend, he uses his cellphone to associate a note to Jules reminding him to do so. On Friday he is casually lunching at a dinner when Jules walks in and John receives a notification that tell him Jules is near and what he wanted to discuss with him.*

In this scenario the application allows the user to associate the notification with a person, this can be a note like the one on the example for the person itself, a reminder, or it can be a notification to be sent to the second user when they are both near each other. This case does not use fixed locations as a mean of detection but the proximity of users.

### 3.8.2.3 – Notification associated with a friendly contact through a third person

*John wants to tell Alice that Jules is an expert at cooking, he sets up a notification on Jules that is intended for Alice. Later that day Jules meets Alice for coffee and Alice receives a notification from John informing her of Jules cooking abilities.*

This is an interesting feature that allows users to tag people with notifications much like taking post-it and putting it on someone's back. That user is then a carrier for the notification for others.

### 3.8.2.4 – Recognizing people in the surroundings and their information

*John is starting his job at a new company and has a staff meeting this morning. Once he reaches the room he picks up his cellphone and has a list of people that are in the room and has direct access to some information regarding them. Their Facebook, twitter, personal websites messages.*

This shows how the application can be used to convey personal information regarding users around the blind person. It makes it easier to access the data concerning those users when they are in proximity.

### 3.8.2.5 – Navigation notifications related to location

*John goes to visit a new shopping mall that just opened, once he get off the bus he reaches for his cell, using the application he is given some notifications regarding the surroundings of his current position where is the entrance of the mall and how to get there. He proceeds to the entrance and receives another notification that conveys to him information about the mall, where are the elevators, where are the stores he wants.*

This scenario shows how notifications can be used to help the blind user. Not all notifications are meant to be made by the user itself, or for a specific person. These helpful notifications in public locations can be made by anyone so that anyone that comes into the new environment is able to take advantage of them. In this case its orientation or guiding information but it could be about objects surrounding the user in different scenarios other than a mall.

### 3.8.2.6 – Notification associated with time, space and users (all at once)

*John needs to be reminded later that night at Jules party to talk to Alice about the plans they had for the party. He sets up a notification on his cellphone to Alice after 10pm when he is at Jules house. Alice arrives at 9pm, and walks about, and leaves again to pick up some drinks. At 10:30 she gets back and John receives the notification that he needs to talk to Alice.*

This is the ultimate scenario where the user is able to control everything about how he sets up the notifications. From the time where it is supposed to go off, who must be present and at what place. This allows for great flexibility in triggering the notifications.

## 3.9 – Conclusions

While some information can be made accessible to blind people through mainstream technologies (e.g. screen readers), there are few projects that enable them to receive information beyond the explicitly acquired through audio or touch. This is a severe limitation that endangers the understanding of the surrounding environment and the competence in social arenas. To the loss of a blind person, there are lacking systems that can offer him knowledge about who is around him, who is he passing by, who left the room, which store is nearby or what is written in the news board at work.

We shed light about the limitations, needs and desires of blind people pertaining awareness of the surrounding environments. Particularly, we put a focus on social environments seeking to provide more information about people around the target users.

We envisioned an application taking into account all our research done in the first half of the project that could explore the issues mentioned by our users and assess the usefulness of these tools and how to implement these tools in the best way possible.

# Chapter 4 – System Implementation

Taking into account the results we gathered from the first practical evaluation made of the first prototype and the data we collected from the questionnaires and interviews made on the first round of interaction with the blind users we set the goal to create a better and more practical prototype which would evaluate and explore the new possibilities found and the needs that appeared from the first set of results.

Some of the changes stemmed from usability issues, a few changes were made to the navigation inside the application also the feedback had some minor adjustments made to make it more understandable and less annoying. We considered some suggestions made by the users to improve their interaction with the tool to turn it into something they would like to use on a daily bases and no have to think about it being a hassle to understand.

This second prototype still had emphasis on the people and identifying them and their location, providing some more options on how and when this process takes place by taking advantage of privacy settings and providing the users with the ability to customize their tool. However the main addition in this phase was the usage of notes (notifications) that users could create, tag and share.

The system we designed is meant to be used as a tool for the blind user to be able to create, share and receive information about people, locations and objects. With this in mind the system obviously has accessibility implementation concepts in order to be easily used by visually impaired people. We take concern on the method of data input, data output and notification method. We thought about building the system piece by piece adding functionalities on each iteration.

The main features of the application focused on the social component the people and the note system (notifications). We tried to cover all the scenarios postulated in our previous step not only those we actually envisioned but also those that came through the experience of our users and the results of their usage.

The technical aspects of the components used in the application will try and be summarized in this section. From the database structure, to how the prototypes were designed and how they worked, every aspect they compromised from feedback to navigation and interaction. We will also do a summary of usage of the application.

## 4.1– System Architecture

The system worked around a relational database on which we stored important data such as backup of the contact list of our users, information about the creates notes and also the important ids from the Google cloud messaging system which was used to relay information back to the users by push technology instead of pull.

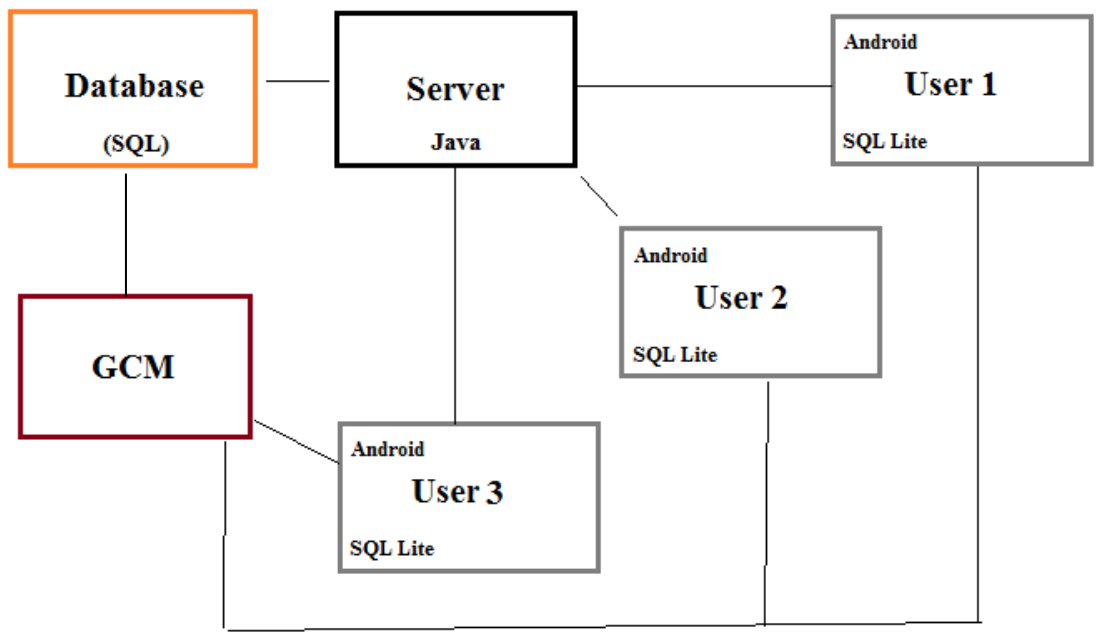

**Figure 12 System Architecture**

The figure explains the overall architecture of the system. Our users were given the Android smartphones running our application developed for the Android devices with their personal database used to store their personal information. The server running on a Java implementation was used primarily to provide a bridge between users. A simple server which the devices running the application would communicate to whenever they wanted to transmit some data. Usually that happened either by a need to share a note with other user(s) or privacy details.

We used the Google Cloud Messaging system to send data to the application whenever needed, again most cases were either note sharing or privacy settings.

### 4.1.1 Google Cloud Messaging (GCM)

Following the tutorial provided by Google itself we setup a developer account on the Google website which allowed us to have an id which we would use to register devices onto the server. Upon registering a device on the GCM an id would be returned so that whenever we wanted to send a message to that particular user we knew how he is identified by the GCM. That data was stored on our database and had a direct relation with our user identification. This system allowed us to have a push message system so that our application did not have to constantly have a connection up to receive messages.

### 4.1.2 – System Database

There were as overall design for a database and this design was implemented on a local database for the devices and on the server side. Although the design was the same, the implementations were not the same on both places. We did not require the server side to have all the data the client had so not all of the tables were replicated on the server side. Also some parts of the server did not need and should not be replicated on the client side so there was also data on the server database that was not corresponded on the client.

The local database was primarily used to store information about the contact associations made on the application between a contact in the user phone and a MAC address detected by the smartphone and to store information about the created notes and who they belonged to. There was another use for the database and that was to store the history of usage of the application, commands done, menus navigated, addresses found. This was however done for the purpose of the tests and not for the functioning of the prototype.

The local database was very simple compromised of eight (8) tables. tables, 6 of them for the functioning of the application and 2 for the purpose of storing history and details of usage.

There was a table (ListaMACs) that contained information about the contacts and the associated MAC addresses. The contacts on this table were identified by the number id that the phone provided them with on the contact list. We used that id to connect the contact directly from the user phone book.

The history tables were divided into two tables, one stored the devices (MAC addresses to be exact) found while the application was active and if they were already associated with a contact they stored that information as well. The second history table was used to record every action taken by the user, every command the user made every screen they went to was registered on the database.

These history tables could at any point be converted to a text file with all this information in a more clear way.

We decided to have just one table for all the notes with different data fields to simply identify the type of note and the data in each of the notes. This table allowed for notes to be created as simple notes not attached to any contact.

Since the final system allowed the notes to be associated with one or several places as well as one or several locations two new tables were created that simply allowed to associate people (contacts) or locations to the notes. Always referenced by a note identification these tables stored information about the notes.

For the location specific case we also created a table to store locations and provide them with an identification. This table stored indoor locations and also outdoors locations that the users might wanted to identify. One of them used MAC Addresses of the devices used to pinpoint the indoor location as data while the other one used the GPS coordinates of the location as data.

Also on the notes another functionality added on this prototype was the fact that these notes could be sent and received between users. For that we made a table that stored information about these received notes since there could be some variations from a created note by the user. This table stored information about the received note. There was also a table created much like the people and locations table that stored information about the receivers of the notes, this table stored associations of note ids with contact ids so that we could store who did the users sent the notes to.

Finally another feature of the final system were some privacy settings and some configuration settings. For this the database needed to have two tables on for blocked users and another for ignored users. Since the application reacted different to both. One was used to store which users the application didn't want to be aware of their presence (ignored) and the ones which the application didn't want to make itself aware to (blocked). Most these were identified by their MAC Address in some cases the id number was used instead.

The server database stored some information that was already being stored in the users' local database. We decided not to make a simple copy of everything that was done locally instead the database was molded to contain only the needed information for the server to do its functionality.

The basic needs were the notes that were being sent between users, those needed to be stored. Information about blocked users, since the only way that another application could know they were being blocked is to through the server. Also since we were using a push system in conjunction with Google Cloud Messaging we needed a table to store the registration ids of the devices so that we could establish the communication between devices.

The tables for the notes and the ignored users are pretty similar to what was made on the local database in order to make it easier to maintain some integrity (see further) of the information.

The most important table that stored the registration ids was associated with the MAC address of each device which was the only way to uniquely identify each one. On this database the identification that corresponded to the Reg_Id table was the one used to identify each user in the other tables.

## 4.2 - Communication and synchronization

This was a problem we had to face especially since the applications objective is to be fully functional even when the user does not have access to an internet connection. For the purpose of this prototype we managed to guarantee a 100% uptime of internet connection which meant we had no need for this to be implemented. However there were plans to implement synchronization of the data made between the local database and the server database.

For this the plan was simple, every change made locally that would need to be updated on the server always checked for a connection and tried to send the information to be updated on the server. In case of failure or that there was no internet connection the tables that needed had a column with a simple tag that informed us if the data was altered or recently created.

Upon establish a connection the application would trigger itself an upload process to the server to synchronize all the data.

As far as the communication went we used the devices 3G capabilities to have wireless connection to our database whenever there was a need to receive updates from a push system or the need from the users to share something to our server. The image describes the flow of information between the user and our communication components.

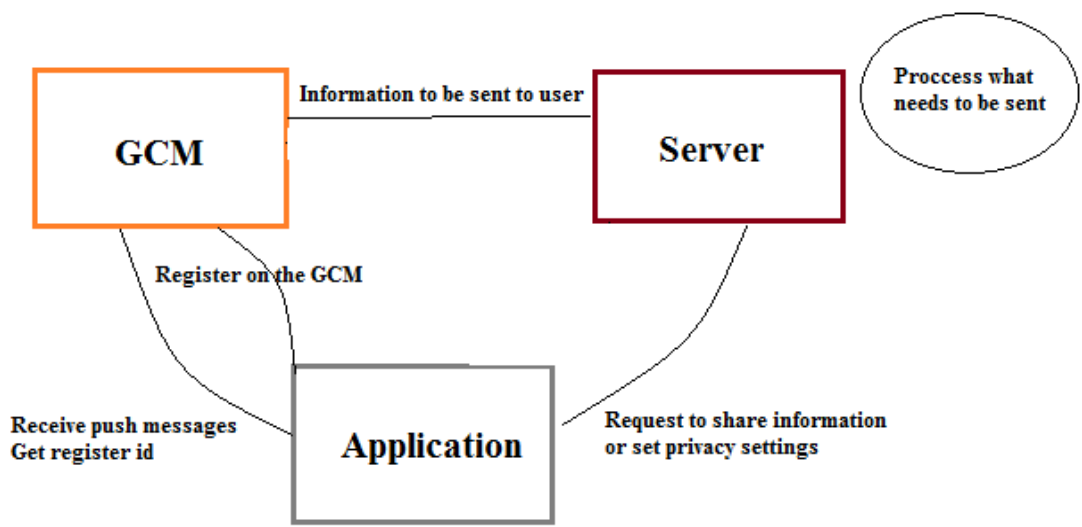

**Figure 13 - Communication overview**

The application registered one time only on the GCM server and received an id from that registration. That id was then sent to the Server so that the server always knows how to use the GCM to relay messages.

After that the communication is done from application to the server, requests to share notes, receive information, or any other request would be sent to the server. The server processed this request or update from the application and if necessary relayed to the

GCM what users needed to be sent information. The users would only get information from the GCM push server.

## 4.3 – User Interface

Our interface was designed was designed orientated for the blind user. This meant that most of the visible part of the interface was mainly for developing purposes since our users did not use this part of the interface.
We used gestures to navigate through our application and sound to provide most of the feedback in the form of text to speech. For the input part of the application we either had sound recording for the notes or we adapted an existing project which allowed for text input for blind users.

### 4.3.1 - Navigation

The navigation we decided to use on this application was through the use of gestures using the GestureFilter from the Android where we identified a few simple gestures. There were some changes between the first and the second prototypes however the main scheme remained the same.

There were six (6) major gestures that the user could make in each screen to trigger a response from it, four swipe gestures, up down left and right, a double tap and a long press. The swipe gestures were mainly used to navigate through each option on the screen, being that the left swipe would almost always be a return command to the previous screen. The remaining commands, right swipe, double tap and long press varied from screen to screen although also in a majority of cases the double tap was used as a confirmation/insertion gesture and the long press as a request for aid/information gesture.

In some cases there was the need to be more specific with the gestures and so we had to distinguish between locations of long presses so that we could had more possibilities of interaction to one screen.

### 4.3.2 Creating a note (Input Writing text)

We developed two ways of inserting text in our application even though one of them was not experienced with much (first prototype) while the second one was the most used. We had to do this since the Android devices we were using (2.3) did not possess accessibility features good enough that the users could use the smartphones keyboards to input text properly.

The first method created involved creating a circle paradigm where the alphabet would be organized as a circle and the user just had to do rotation gestures to navigate through the alphabet and press once to insert the desired letter.

The second method was already created and designed in previous projects [7] and showed great success hence why on the second prototype it was used as the method to insert text in the application.

There were some changes made to the functioning however since we needed some more commands to be detected. The users had the alphabet rearranged in a matrix grid where they could navigate up and down through vowels and left and right through the normal order of the alphabet. In this particular case we had to detect different types of long presses in order to distinguish between erasing a letter or leaving the screen or asking for information or simply confirming the insertion of the text.

### 4.3.3- Feedback

The feedback provided in the application was done in two ways, first through vibration that was mostly used to inform the users of the existence of new notifications and the audio which was primarily used while navigating through the application only.

#### 4.3.3.1 -Vibration

For the vibration we tried to setup some simple patterns to distinguish between the different types of notes that the user was being warned about. Trying to create a code of some sort that allowed users to not require the phone to be taken of their pockets to realize what were they being notified about.

The three patterns we decided to implement were to distinguish between a warning about a person being near, an audio note that was triggered and a textual note that was triggered.

- Long 500ms vibration for a person being near
- Shorter 250ms vibration for an audio note
- Pattern of 50ms vibration -> Stop -> 250ms vibration for a textual note

We later wanted to implement a lot more patterns to be recognized as the second prototype had a lot more features that could use with some unique identification.

### 4.3.3.2- Audio

The audio feedback was used over the Text to speech to navigate through the application and whenever requested provide the user with audio information about their location inside the application and what was possible for them to accomplish in the screen they found themselves in.

Most important was the feedback provided as actions were completed in order for the user to be aware that something was accomplished with success.

We wanted to add different types of audio feedback later, some tones for the notifications or even music that the users could associate with their notifications.

### 4.3.4- Menus

This is a simple description of how the menus were implemented.

In the navigation screens the user had the option to scroll through a list of options, in each he had a choice to enter it and that would lead him to a new screen. When finishing a major operation the user would be returned to the main screen most of the times so that he could always be situated.

We fiddled a lot with the arrangement of the options, some we thought had more priorities and some options we tried to figure the best place inside the overall scheme of the application of where to put them.

# Chapter 5 –Evaluation

## 5.1 - Introduction

The tool expanded its options and added some main features that had the objective of providing the blind user with more information on their environment as well as providing the blind user with the ability to interact with the environment and other people. We took our first experience from the first evaluation we made in order to improve this method of testing out our application.

We give a brief explanation on the main points we had to focus on figuring if we had successfully corrected them.

One of the issues we identified after the first evaluation made was the lack of options and customization by the user. The users wanted ways to control the tool and be able to select when it would work or not and how it would work, for that we implemented some options. Some focused on privacy settings they could adjust, two main features where blocked and ignored lists, the blocked list was something they could use whenever they wanted a specific user to not be able to take notice of their presence, anyone blocked would not be able to detect the person who blocked. The ignored list on the other hand was used when the user wanted to purposely stop identifying or recognizing one person, they could ignore that user for whatever amount of time they deemed necessary and that person would be ignored by the application.

It was also added the generic mode of invisibility where the user could simply make himself invisible and not broadcast its presence therefore making other users unable to detect their presence.

Another usability concern was the fact that they could not prevent the application from keep giving them warnings on unwelcome timings therefore we implemented a feature that allowed the users to silence the application. Although the application would still maintain its normal functions it would no longer disturb the user with warnings unless the feature would be toggled off.

A big part of this final implementation was the ability to create notes and use those notes in a multitude of ways.

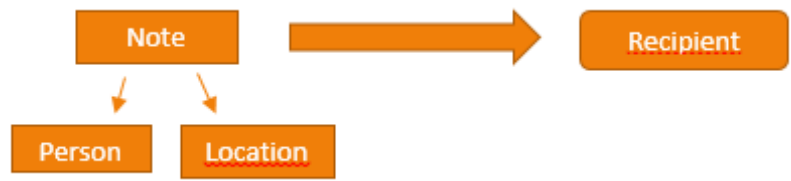

**Figure 14 - Scheme of note interaction**

When creating a note the user could choose between an audio (voice recorded) note or a textual note, once that note was created it was simply used as something that the user could check at any given time if he so chose to. Once the note was created (audio or text) the user had options to add some triggers to the note. What we call triggers could be people that were previously added to the contact list of the application or locations, either indoor or outdoor. These triggers are what make the notes be shown to the user when those events are registered. In case of a person associated it meant that the note would be shown to the user when that person was detected in the proximity same goes for the locations whenever the user was near one of the locations associated with a note that note would be shown.

The outdoor locations could be added when the user was on the spot they wanted to associate or from a list of previously saved locations, a list that the users could fill up as they could save outdoor locations whenever they felt like it, all they had to do is save the location when they are there and from that point on the location is always available to be associated (added) to a note.

For indoor locations we used some devices to tag some key places inside the area in which the users spent most of the time (the foundation), those could not be added manually but were already present.

Finally the recipient of the notes made it so that the note was not only shown to the user that created. Adding a recipient to a note made it so that users could receive notes created by others users, that way user A could create a note associated with a location and send it to user B, when user B would be in the location specified in the note he would be shown the note created by user A.

The combination of locations and people associated with the notes, as well as the ability to send notes between users allows for a lot of different ways that messages can be used, from reminders to a user, messages from one user to another, public notes anyone could see about a place or a person and so much more.

## 5.2- Evaluation Method

We returned to the foundation where we had done our first round of questionnaires and tests in order to have our second prototype evaluated. We decided to let our volunteers (blind users) try out the application for an extended period of time so we could get more data out of it.

### 5.2.1 - Material and applications

For this evaluation we used three smartphones where the application was deployed. Two Samsung Galaxy Ace and one Samsung Galaxy Mini running Android version 2.3 on all three devices. It was also used 3 other low end cellphones which had Bluetooth capabilities with the purpose of providing an indoor beacon to tag locations. For the devices used for being detected we used the users and the participants personal devices all of them Nokia's cellphones with Bluetooth capabilities.

The application deployed was built upon the previous prototype and enhanced to provide the latest features.

### 5.2.2 – Participants

We had 4 (four) participants in this trial, young participants and older ones being over 50 all spread across the duration of the trial. Three of them had already been a part of the previous probe that was done so they were already familiarized with the work and the application. There was one participant that had only participated on the questionnaire that was realized on the first evaluation but did not had any contact with the application at that time.

### 5.2.3 - Procedures

The trial ran from a total span of 7 days from around 9/10 am to 3/4 pm. At the start of each day the low end cellphones that were used to tag indoor locations were setup and then the participants of that day were given the Android devices with the application and if there had been a modification done to the functioning of the application from the previous day they were informed about it.

The table that follows describes which users were trialing the application throughout the 7 day duration.

| User/Day | Monday(1) | Tuesday(2) | Wednesday(3) | Thursday(4) | Friday(5) | Monday(6) | Tuesday(7) |
|---|---|---|---|---|---|---|---|
| User 1 | X |  | X | X | X |  | X |
| User 2 | X |  | X | X | X | X | X |
| User 3 | X |  | X | X | X |  |  |
| User 4 |  |  |  |  |  | X | X |

**Table 6 - User testing schedule**

The first week was realized with the same 3 users having Tuesday with no one due to some critical issues that had to be corrected from the first day of usage.

The users started to build up their usage of the features implemented in the application from day 1 they started with some of the basics and simpler things and worked their way up to the most elaborate features like creating notes and sending them or association people and places to the notes.

They were not imposed any guide to follow on things they had to try or features they had to do. At the start of the day we would speak about some of the things they could do and they were allowed to freely explore throughout their day whenever they felt like it.

During the day there was almost always (excluding 1 day) someone present at the foundation in order to fix any problems or to help any user if there was something they were not understanding or had any doubts about.

At the end of each day the devices were collected and a small amount of feedback was taken, things like problems using the application, things that did not work properly or any small comments they might have come up with while using the application.

Every action made by the users and the detection history was logged by each device and extracted at the end of each day or two.

At the end of the 7 days of evaluation, in the exception of 1 user which had to be conducted after 5 days, there was a questionnaire that was realized to all the users where we tried to evaluate more objectively how they felt about the application and more importantly the features that it provided them. Its usefulness, usability, difficulty if they were excited about the possibilities that the tool provided.

## 5.3 - Results

Our prototype evaluation lead to some results gathered from the entire procedure. We had some research questions we wanted to answer and for that we had our evaluation prepared to try and answer those questions. The questionnaires, the debriefings, data logs, final discussion with the users all was aimed at trying to answer these questions.

R.1 – Do blind users have awareness problems about their environments that can be reduced?

R.2 – Do context awareness tools help the user understand more about their surrounding environment compared to what they normally experience?

R.3 – Do users feel comfortable with these tools?

R.4 – Would users like to have these tools on their daily life?

### 5.3.1 – Explored and unexplored scenarios

We had our list of scenarios identified that we aimed to cover with our implementation of the application. Some of these were explored on our evaluation made with the users, other were not. Some more than others. We also had scenarios not explicitly specified that were explored on this evaluation.

Associating notes and reminders with contacts was explored by our users. After having the application setup and the previously needed steps completed such as associating contacts and finding people known to the user, our users explored this by creating notes and having them associated with people they knew.

One of the most explored and experienced scenarios was the recognizing of people and conveying information about those people. The application was constantly performing this to our users.

Some scenarios were not explored much or at all. For instance there was never a three way situation where a note would have association of a time a place and specific users to be triggered. However some scenarios came up, like providing information regarding a proper place, where specific places had specific notes associated.

### 5.3.2 - Opinions and quotes

Users were enthusiastic with the possibilities the system provided. However they do pointed out some flaws. They pointed out the system would be more useful outdoors, one of them even said that "It would be much more useful outdoors then indoors", since our evaluation was mainly done always indoors they did not get to experiment with the possibilities it could have been opened with the outdoors environment.

One use that was not implemented that was mentioned by one of our users was for instance "Transport companies, so they could give information about the buses times and places", a way to have on location access to information regarding transportation. This was also an outdoor feature that could be very useful.

Someone mentioned that they should be able to configure the information they would get from notifications. We did not provide this since our notifications were all shown the same way, and just depended on whether they were notes or proximity notifications. It is something that could be added for more adaptability to each user.

There were also some comments made in regards to the feedback, that "It could vibrate but also provide with a sound warning" which was something that we also not considered since we thought that it would just overload or cause too much constant harass to the user.

### 5.3.3 – Data analysis

First of all we had data from specific usage of the application throughout the duration of the evaluation. All 7 days had log files with detailed information about each of the users (on the days they were using the application). This data allowed us to analyze everything from the screens visited, the options used, how they interacted with every component of the application.

We had also logs pertaining the application detection functioning, we could sort and try to find normal patterns of detection and try and connect what the users were doing to what was being detected around them (people mostly).

We had at the end of each day a small talk to try and correct any issues that were found on that day that could easily be improved on and could have some impact on the evaluation this method was used throughout the entire evaluation.

Finally we had the most important part of the evaluation where we tried to access through interviews (questionnaire) how did the application faired in the eyes of the users, what they liked what they didn't and every piece of information we could get from them.

The questionnaires were recorded so we could also get some contextual information besides the direct answers from the questionnaires.

### 5.3.3.1 – Analysis of usage

We analyzed the logs of usage of the application and we tried to find what were the most used options and the most used screens. Trying to assess what the users found easy and useful to use and what they found hard or not important in the application.

- Request for instructions

    This was one of the features that was request by all the users and even though one may think it was the most used it was not the case. It had a fairly medium usage compared to all the other features. It was used about 10-20 times a day per user which I think it's a fairly low number for an application with so many options and with users with fairly little usage of both smartphones and touchscreens.

- See notes

    This was indeed the most used or visited screen of the application, mainly because this allowed for the use of many other features since the note menu was the access point. Users tended to send notification to people or associated people to the notifications trough this screen, these two were the main

- Create notes

    This was side by side with the see notes screen the most access screen/feature.

- See notifications

    This feature was very commonly used as it was the easiest access to get information from the application. And it was the main concept of the system. Relay information to the user.

- Send note

    This feature was used a good amount of times to send out the notes. Mostly used to send notes to different users and to setup or send a note to a specific location. It was a feature that allowed more freedom to users to share their information therefore it was important to verify that they took advantage of it.

- See people near

    Even though the notification system already showed this information per case, this feature allowed for the user to be able to get all the information about people near at once. This was used a few times, but mostly when wanting to add another contact to the list instead of using it just as a source of information about who is around.

**5.3.3.2– Detection Results Logs**

The log files provided us with an easy read if it was a detection of someone arriving or if it was a device disappearing from range (i.e. someone moving away), we had the time and the coordinates of the detection, however these coordinates were "fake" given that it was mostly always indoors and so we could not analyze any location patterns.

Example of log:

```
Saiu   desconhecido   -   34:C8:03:F6:F3:A8        Time:   25/07/2013
11:02:57.000     Coord: 38.738522;-9.1543572
Entrou  desconhecido   -   Jj  34:C8:03:F6:F3:A8    Time:   25/07/2013
11:03:04.000     Coord: 38.738522;-9.1543572
```

Throughout the evaluation these logs are very hard to analyze, it's hard to find some patterns but considering we knew the location was restricted and we knew the scheduled we could extrapolate easily in which conditions did the application registered the most number of detections of unknown people as well as known people.

First off the class times and the lunch period was the one where we had the most number of devices detected, things settled down on the hours where there was less movement and the users stayed stationary.

Most devices detected were always unknown devices that were not added to the application, the users could freely add and explore at their own will which they did on some occasions but in a very small percentage of cases.

These logs can be used to perform some pattern analysis and try and make some correlation between the users actions and what was surrounding them at the time, which people, if known or unknown, a lot of people or very few.

In that case we noticed that the users were more active in the application usually when small groups of people were detected and not in the times were several devices were being detected at the same time.

Of course we also noticed that most of the use was made while there were known devices around, given they knew the evaluation of the method was being done that could have influenced the fact that they felt compelled to "play" with the application when some fellow users were around.

Overall there were detected close to 100 different devices throughout the experiment even though we had some people warned that they could leave their Bluetooth on discoverable mode these results show that there are already a lot of people that have this option by default allowing us to detect them allowed us to identify the times with more movement.

## 5.4 Questionnaires

After our evaluation ended we made individual questionnaires to each of the participants that participated in the week and a half of its duration. The questionnaires had a more objective approach and most of the questions were made in "like" scales in order for us to be able to quantify some data about satisfaction and performance of the system.

However there were also some open questions for some input on how the system performed to allow for the users to freely give their opinion.

This section presents the results of the questionnaires regarding the objective questions made to our users.

These objective question had a scale from 1 to 5, 1 always being the worst value and 5 the best value.

5.4.1 – Note Creation: Q. Did you had any difficulties creating textual notes?

We had two of our users claim that creating a note was fairly easy while two others claiming it had a bit of difficulty. This was mostly due to the method used to insert text. Some had previous experience and others did not.

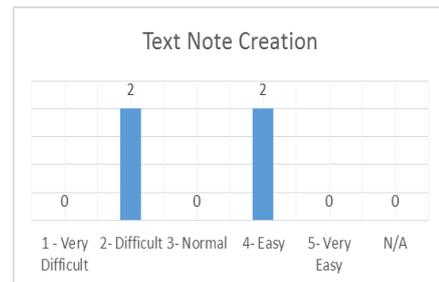

**Graph 1 – Text note creation**

5.4.2 – Note Creation: Q. Did you had any difficulties creating audio notes?

Again the answers here were spread across the scale, and one of the users did not had the chance to try Audio notes.

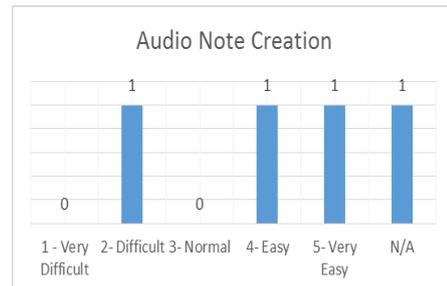

**Graph 2 - Audio note creation**

5.4.3 – Device Association: Q. Any difficulty associating a device to a contact on your list?

All the users reported no issues associating a device to a contact on their list.

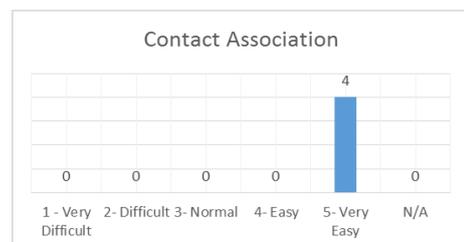

**Graph 3 - Contact Association**

### 5.4.4 – Note Association: Q. Had any difficulty associating notes with people?

Our user did not present any major difficulties when trying to associate people to the notes they created. Although some seemed more comfortable than others.

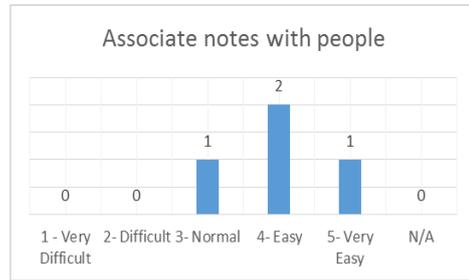

**Graph 4 - Associate notes with people**

### 5.4.5 – Note Association: Q. Did you find useful being able to associate notes to people?

Users found this feature to be very useful to have in the system.

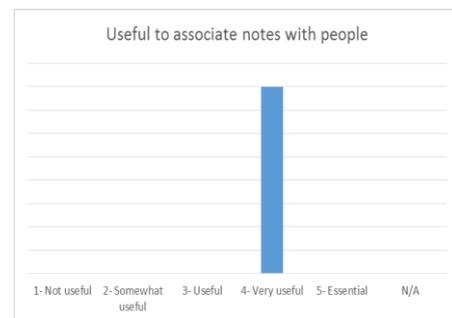

**Graph 5 - Useful to associate notes with people**

### 5.4.6 – Note Association: Q. Had any difficulty associating notes with locations?

Two of our users did not use the feature to add a note to a location therefore they had no answer in this question. The other two find it fairly easy to achieve.

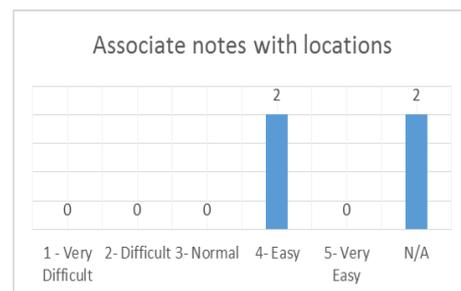

**Graph 6 - Associate notes with location**

5.4.7 – Note Association: Q. Did you find useful being able to associate notes to locations?

Our users found it useful to be able to associate notes to locations. Even those who did not get the chance to try the feature answered positively on the benefits of it.

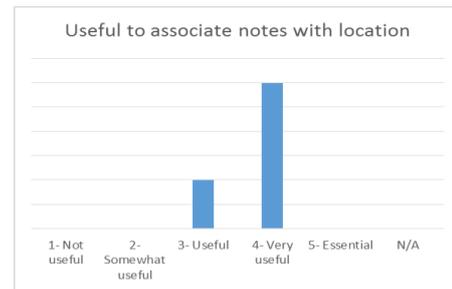

**Graph 7- Useful to associate notes to location**

5.4.8 – Sending notes: Q. Had any difficulty sending notes to users?

This was on part of the system that had the most problems and was reflected on our questionnaires. One of the users did not manage to send a note and the ones that did found the process not so easy.

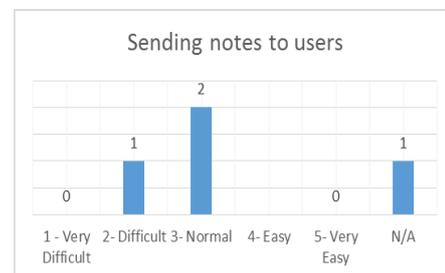

**Graph 8 - Sending notes to users**

5.4.9 – Notifications – Q. Were the notifications easy to understand?

Users found the notifications simple to understand and intuitive. There was one of the users that thought the lack of information on them was a bit too much.

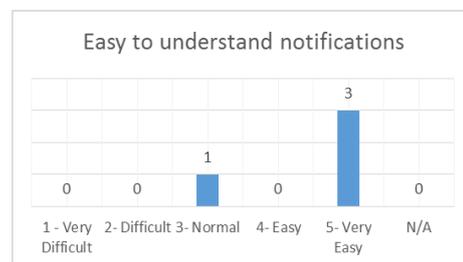

**Graph 9 - Easy to understand notifications**

5.4.10 – Notifications – Q. Managing the notifications was easily achieved?

Users reported no problems with deleting and browsing through the notifications they had. It was the easiest part of the system to handle by their standards.

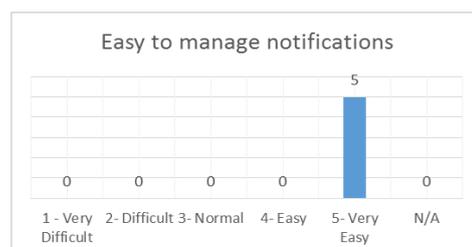

**Graph 10 - Easy to manage**

## 5.4.11 – Ignored List: Q. Did you find useful to be able to ignore people?

Even though users did not try out this feature the question regarding whether or not it was useful was a positive one. All users found the feature an important part of the system.

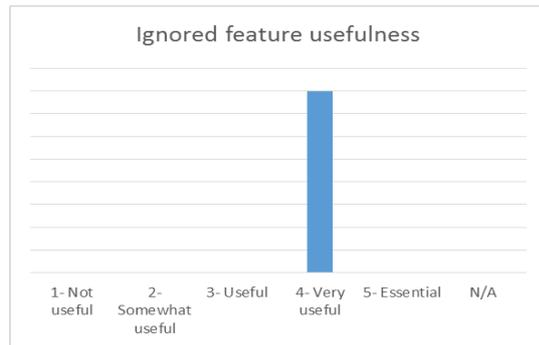

**Graph 11- Ignored feature usefulness**

## 5.4.12 – Block List: Q. Did you find useful to be able to block people?

Same thing as with the previous question.

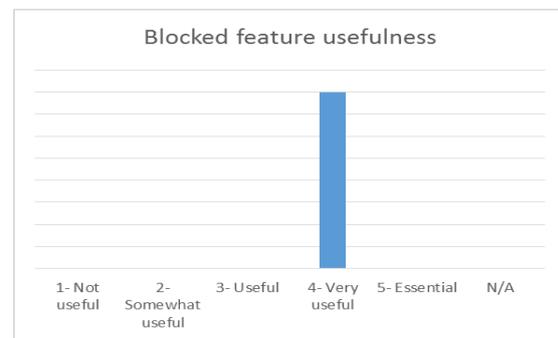

**Graph 12 - Blocked feature usefulness**

## 5.4.13 – Silence Mode – Q. Did you find useful to be able to have the application in silent mode?

Again same result as the last question.

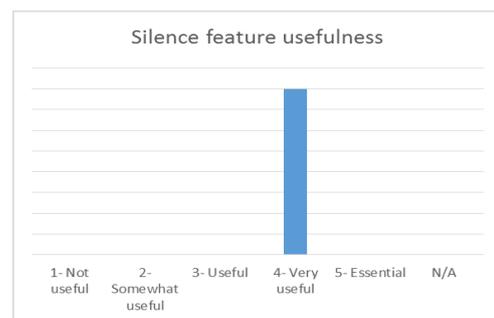

**Graph 13 - Silence feature usefulness**

## 5.4.14 – Feedback – Q. How important you consider the feedback provided was?

With no surprise the users considered that the feedback was essential to the system. Without it they would get lost.

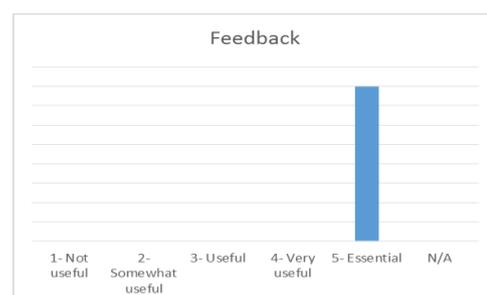

**Graph 14 - Feedback**

## 5.4.15 – Q. Would you be interested in having this application to be used at your disposal?

Overall our users were satisfied with the possibilities of the system and when asked if they would like to have the system available for them the response was positive.

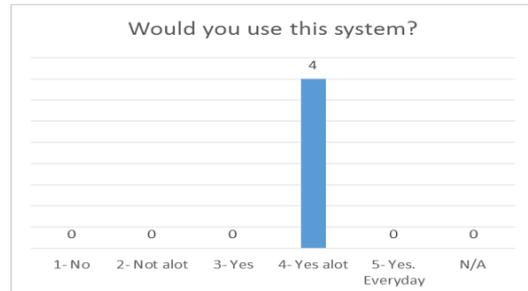
Graph 15 - Would you use this system?

Other questions that were also made let us know that the users found it useful to be able to have both audio and text notes as alternatives to create notes. One complements the other in different situations. There was some feedback in regards to the note

## 5.5 Discussion

We believe that our project allowed blind users to experience a taste of what awareness tools can do for them. We had some success gathering positive feedback and good information into what exactly helps a blind user and what should be implemented on an awareness tool. Not all of the fields have been explored since identifying people and having a notification system can be just a part of a bigger better and more complete system. Even if these are two important components of awareness that blind users lack the most.

Even so our aim was to provide the first step into the design of an awareness tool and reach out to a user oriented approach to get from the target population the best feedback possible.

The research part of our project went extremely well as well as our first prototype approach which allowed the full system to be much more adapted to our users which allowed for good results with the complete system.

We believe blind users can benefit from these tools on a daily basis but they need to be developed with the proper focus to the intended users.

The technology used is secondary as it could be developed on an iPhone using Wi-Fi instead of an Android with Bluetooth. The technology that would fit the best would be the one who could support the accessibility features the best. Since the Android versions we worked on had o accessibility features even the text-to-speech was not a standard application and had to be paid and downloaded.

We would have liked for a more extend period of testing which we could have simply left the devices over night with the users and have a less controlled experiment but that was not possible. That sort of approach could lead to better scenarios and adaptations for the system.

The difficulty for our users in using touch technology was non apparent, as it took them almost no time to get used to it.

Our research question number one, "*Do blind users have awareness problems about their environments that can be reduced*", was answered through our questionnaires made at the start of the project where we managed to identify the issues blind users had about context awareness in their surroundings. Their discomforts and difficulties were made aware with that first questionnaire where every interviewed person identified some sort of difficulty because of their condition that involved the lack of information about their surroundings.

Research question number two, "*Do context awareness tools help the user understand more about their surrounding environment compared to what they normally experience*", was answered with the development of our system and its evaluation and also with the help of our research made on the area. Our system managed to provide an increase in the amount of information that a blind user gets when they are in an environment. The application showed that allowing the users to recognize people near them and increase the information in their surroundings through the sharing of notes allow them to have more information to work with then they usually do.

Research question number three, "*Do users feel comfortable with these tools*", was also answered through the evaluation of our system. The results were positive, no user had issues with the utilization of the system. They managed to perform most of the actions with no problems after a small learning curve. Of course the system did had some flaws which were pointed out by our users during the evaluation, some interaction design mistakes which made it harder sometimes for them to work with the application.

Finally our research question number four, *"Would users like to have these tools on their daily life",* was our final question to our the users we had perform the evaluation. After a week and a half of usage and interaction with the system they had the ability to gage the potential of the system and what benefits it could bring them to have this sort of tool available to them. All of our users answered positively in this subject confirming that these tools would be a benefit for them and that they would enjoy having them as auxiliary awareness tools.

# 6 - Conclusions

We have established that lack of awareness is indeed a fact that is present in the day to day life of blind users and more importantly that they are open and willing to have tools to help them improve this awareness.

We showed the most common origins of discomfort from awareness which are the lack of knowledge of people in the environment and not being able to perceive certain elements of the environment can be provided to blind users through the aid of current day technology such as smartphones.

The system developed demonstrates how it is possible to have an awareness tool that is accessible to blind users and can provide, with the aid of its technology components, the awareness that blind users lack.

Even though we think the overall project was a success, we had some issues on our final evaluation. Not everything was able to be properly tested by our users. Even though we had a larger period of testing there was the need for a longer and perhaps more oriented evaluation.

Awareness tools can help provide a new perspective on how environment is perceived by the blind user and make them able to have a more sociable and intractable life.

This field is yet to be fully explored and there is a possibility for a development of a bigger and larger system which can be used with the same basis of sharing information between users.

## 6.1 – Limitations

We wanted to implement a much larger system then the one we were able to. The awareness tools can exist in a multitude of forms and not just identifying people or being able to tag/share information. There we many awareness issues that can still be explored and implemented to help enrich a full awareness system.

Our evaluation had a more controlled approach mostly in indoor environments which presented a involuntary constraint on scenarios and also the free thinking of our users. As well as the time period used to do the evaluations, taking the devices back at the end of the day also cut down some possible scenarios that could have been explored otherwise.

A system designed to work with all the capabilities of a smartphone can be extremely battery exhausting. Requiring many of the devices technology and also an internet connection would make this system not so globally available for any blind user.

One other aspect to be explored which we did not do is the balance between too much information and too little. Users would like to know more about their surrounding and be aware of it, but to much information can lead to overload. That balance to find the correct way of providing information is needed.

## 6.2 – Future Work

Following up this project an assessment of the difficulties encountered by our users in the evaluation phase would have to be adjusted and corrected to improve interaction and simplify some actions that caused more trouble to our users.

An implementation on a broader scale with more users and a larger test timeline would also be important to be able to explore every scenario multiple times and allow our users to discover new possible scenarios.

Some areas that could be explored in context awareness that we did not do are object recognition, obstacle recognition, space visualization or even distance. Awareness tools can be of great assistance to blind users and there are still plenty of opportunities waiting to be explored.